%% file: draft_2.tex
\documentclass[pra,twocolumn,superscriptaddress,showpacs,amsmath,amstex,amssymb,citeautoscript]{revtex4-1}

\newcommand{\vect}[1]{\boldsymbol{#1}}

\usepackage[T1]{fontenc}
\usepackage[utf8]{inputenc}
\usepackage{lipsum}
\usepackage{amsmath}
\usepackage{amssymb}
\usepackage{bbm}
\usepackage{braket}
\usepackage{xcolor}
\usepackage{pifont}
\usepackage[mathscr]{euscript}
\usepackage[shortlabels]{enumitem}
\usepackage[justification=justified]{subcaption}
\usepackage{graphicx}
\usepackage{lipsum}
\allowdisplaybreaks
\usepackage{float}
\usepackage{mathtools}
\usepackage{graphicx}
\usepackage{dsfont}
\usepackage{comment}
\usepackage[colorlinks=true]{hyperref}  
\hypersetup{
    bookmarks=true,         % show bookmarks bar?
    unicode=false,          % non-Latin characters 
    pdftoolbar=true,        % show Acrobat
    pdfmenubar=true,        % show Acrobat 
    pdffitwindow=false,     % window fit to page when opened
    pdfstartview={FitH},    % fits the width of the page to the window
    pdftitle={Band theory for heterostructures with interface superlattices},    % title
    pdfauthor={Putzer, Pupim, Scheurer},     % author
    pdfsubject={},   % subject of the document
    pdfcreator={},   % creator of the document
    pdfproducer={}, % producer of the document
    pdfkeywords={} {} {}, % list of keywords
    pdfnewwindow=true,      % links in new window
    colorlinks=true,       % false: boxed links; true: colored links
    linkcolor=blue, %red,          % color of internal links (change box color with linkbordercolor)
    citecolor=blue,        % color of links to bibliography
    filecolor=magenta,      % color of file links
    urlcolor=blue           % color of external links
}

\newcommand{\equref}[1]{Eq.~(\ref{#1})}
\newcommand{\equsref}[2]{Eqs.~(\ref{#1}) and (\ref{#2})}
\newcommand{\secref}[1]{Sec.~\ref{#1}}
\newcommand{\figref}[1]{Fig.~\ref{#1}}
\newcommand{\refcite}[1]{Ref.~\onlinecite{#1}} 
\newcommand{\refscite}[1]{Refs.~\onlinecite{#1}}

\newcommand{\tableref}[1]{Table~\ref{#1}}
\newcommand{\appref}[1]{Appendix~\ref{#1}}

\newcommand{\pdagger}{{\phantom{\dagger}}}
\newcommand{\diff}{\mathrm{d}}

\renewcommand{\approx}{\simeq}

\newcommand{\ie}{i.e.~}

\renewcommand{\vec}[1]{\boldsymbol{#1}}

\definecolor{wrongultramarine}{rgb}{1,0.5,0}

\newcommand{\db}{\delta b}
\newcommand{\vdb}{\delta \vec{b}}
\begin{document}

% not sure about the title. Alternatives/building blocks:
% Oxide interfaces with a twist
% moiré interfaces
%\title{Superlattice engineering at interfaces between different materials}
%\title{Band theory for superlattices at interfaces between different materials}
\title{Band theory for heterostructures with interface superlattices}
%title{Band theory of moiré interfaces}

\author{Bernhard Putzer}
\thanks{These two authors contributed equally.}
\affiliation{Institute for Theoretical Physics III, University of Stuttgart, 70550 Stuttgart, Germany}

\author{Lucas V. Pupim}
\thanks{These two authors contributed equally.}
\affiliation{Institute for Theoretical Physics III, University of Stuttgart, 70550 Stuttgart, Germany}

\author{Mathias S.~Scheurer}
\affiliation{Institute for Theoretical Physics III, University of Stuttgart, 70550 Stuttgart, Germany}

\begin{abstract}
Motivated by recent experiments demonstrating the creation of atomically sharp interfaces between hexagonal sapphire and cubic SrTiO$_3$ with finite twist, we here develop and study a general electronic band theory for this novel class of moiré heterostructures. We take into account the three-dimensional nature of the two crystals, allow for arbitrary combinations of Bravais lattices, finite twist angles, and different locations in momentum space of the low-energy electronic bands of the constituent materials. We analyze the general condition for a well-defined crystalline limit in the interface electron system and classify the associated ``crystalline reference points’’. We discuss this in detail for the example of the two-dimensional lattice planes being square and triangular lattices on the two sides of the interface; this reveals non-trivial reference points at finite twist angle and lattice mismatch, leading to a novel form of magic angles, which we refer to as ``geometric magic angles''. 
We further show that band structures of mixed dimensionality naturally emerge, where quasi-one- and two-dimensional pockets coexist. Explicit computations for different bulk Bloch Hamiltonians yield a collection of interesting features, such as isolated bands localized at interfaces of non-topological insulators, Dirac cones, van Hove singularities, a non-trivial evolution of the band structures with Zeeman-field, and topological interface bands.  
Our work illustrates the potential of these heterostructures and is anticipated to provide the foundation for moiré interface design and for the analysis of correlated physics in these systems. 
\end{abstract}

%\date{\today}
\maketitle
%\tableofcontents

\section{Introduction}
A significant amount of modern technology, for instance, solar cells, semiconducting diodes, lasers, and field-effect transistors, is crucially based on interfaces between different materials \cite{RevModPhys.73.783}. This success is based on enormous research efforts to engineer heterostructures with high precision in which the electrons at the interface exhibit fundamentally interesting and/or technologically desired properties, distinct from those of the involved bulk materials. Apart from silicon-based heterostructures, recent research has also focused on constituent materials with more significant electronic correlations, most notably oxides \cite{ReviewInterfaces2,ReviewInterfaces1}, opening up avenues to interface superconductivity \cite{InterfaceSuperconductivity} and complex magnetic order \cite{magneticorder1,magneticorder2,magneticorder3}.  

In parallel and especially in the last five years, there has been a lot of progress in the field of two-dimensional (2D) materials \cite{Review2DMaterialsLin}, which can also be used for the construction of heterostructures with unique properties of fundamental and practical relevance. A particularly promising and active direction is to construct moiré superlattices from graphene and related 2D van der Waals materials \cite{macdonald2019bilayer,andrei2020graphene,Review2DMaterialsLin,ReviewMoirevdWSun,ReviewMoireTMD,D3TC02660D,MoreMoireReview,NuckollsReview,BerryPhaseReview}, in which the twist angle between the layers constitutes a unique knob to tune the electronic properties. This subfield has gained a lot of momentum as a result of the discovery \cite{Cao2018_correlated, Cao2018_superconductivity} of superconductivity and interaction-induced insulators in twisted bilayer graphene. 

\begin{figure}[b]
   \centering
    \includegraphics[width=1.0\linewidth]{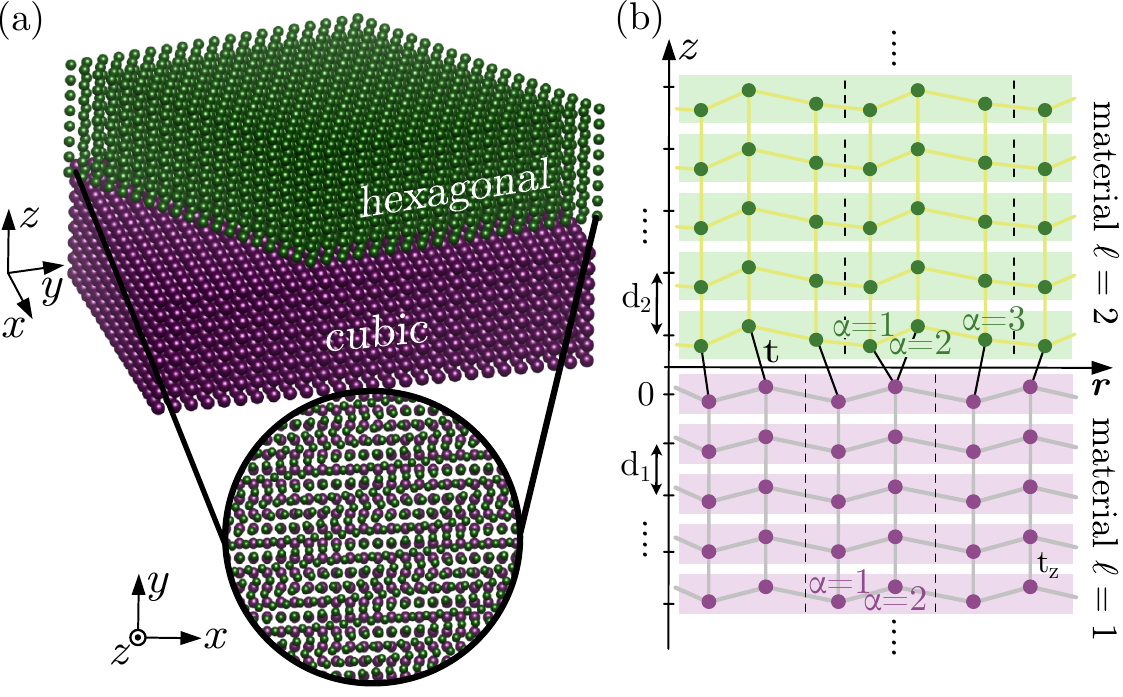}
    \caption{In this work, we consider heterostructures consisting of two different materials with in general different Bravais lattices, shown in (a) \cite{Caption1} for the example of one material having a cubic (purple) and the other a hexagonal lattice (green); this leads to a moir\'e interface system (circular inset). In (b), we show a schematic of a 2D cross section through the system, assuming two (three) Wannier orbitals (centers as dots, labeled with $\alpha$) in material $\ell =1$ ($\ell = 2$) as an example.}
    \label{fig:introfig}
\end{figure} 

While tight-binding models \cite{TightBinding} and first principle techniques \cite{AbInitio1,AbInitio2} have also been developed, a particularly frequently used and insightful tool is the celebrated continuum-model description \cite{dos2007graphene,MeleModel,bistritzer2011moire,dos2012continuum,ContModelBalents}: neglecting relaxation effects, the moir\'e superlattice created by twisting the two honeycomb lattices of graphene by a relative angle $\theta$ only has exact (superlattice) translational symmetry for certain discrete commensurate angles. However, if one is only interested in the electronic properties in the immediate vicinity of the Fermi level (energy range much smaller than the bandwidth), one can expand the dispersion of each graphene layer around the K and K' points, yielding a continuum Dirac theory. Then, for sufficiently small $\theta$ and using that the Fourier transform $T_{\vec{q}}$ of the interlayer tunneling matrix elements decay with large $\vec{q}$ \cite{bistritzer2011moire}, three momenta---related by the three-fold rotational symmetry of the moir\'e lattice---contribute; this creates a triangular lattice reconstruction of the Dirac cones, a non-trivial evolution of the bandwidth with twist angle \cite{dos2007graphene,MeleModel,bistritzer2011moire,dos2012continuum} and non-trivial topological properties \cite{PhysRevX.8.031089,PhysRevB.103.205412} obstructing the construction of tight-binding models for certain subsets of bands.

Sparked by the enormous developments and interest in twisted graphene stacks, related moiré engineering has been studied in a variety of settings, such as twisting the surface Dirac cones of topological insulators (TIs) \cite{PhysRevB.106.075142,KoshinoTwistedTIs}, twisted transition metal dichalcogenides \cite{PhysRevLett.122.086402}, strain control of moiré superlattices properties  \cite{StrainEngineering,TMD_strainengineering}, twisted bilayers of GeSe \cite{RubioGeSe,ToshiGeSe}, BC$_3$ \cite{PhysRevB.107.085127}, FeSe \cite{FeSeTwisted}, $\pi$-flux square lattices \cite{PhysRevB.104.035136,PhysRevB.105.165422}, square-lattice superconductors \cite{twistedCuprates} and Hubbard models \cite{2024PhRvB.109h5104J}, and two identical 2D Bravais lattices \cite{ToshiTwisted2D}, some of which can exhibit emergent quasi-one-dimensional (1D) electronic properties. Furthermore, other moiré systems and related phenomena were also studied, namely: moiré-lattice enhancement of mass-anisotropies \cite{PhysRevB.108.L201120}, twisted magnetic TIs \cite{PhysRevResearch.3.033156,2024arXiv240401912W}, twisted multilayer systems with non-trivial behavior depending on the twist-angle configuration \cite{KhalafKruchkov2019,orderlydisorder,PhysRevB.107.235425}, all the way to twisted graphite \cite{PhysRevB.108.165422} and van der Waals material stacks with Eshelby twist \cite{PhysRevB.103.245206}, which display interesting optical and transport properties. Moiré-superlattice engineering has also become subject of interest in photonics \cite{TwistOptics,PhotonicCrystals,PhotonicCrystals2} and cold atoms \cite{PhysRevA.100.053604}.

Despite the long history of interface engineering at heterostructures involving three-dimensional (3D) materials, atomically sharp moiré superlattices at interfaces have remained challenging due to contamination layers. However, there has been a lot advances in the field in the last few years \cite{pryds2024twisted,twistedoxides,Oxides2,Oxides3}.
Most notably, a recently developed experimental technique allowing for the creation of a contamination-layer-free moiré superlattice with twist angle $\theta \approx 3.5^\circ$ between sapphire (hexagonal) and  SrTiO$_3$ (cubic) \cite{MannhartExperiment}, see \figref{fig:introfig}(a), has opened up a new door towards moiré superlattice design at interfaces between \textit{different} lattices and with tunable twist angle. 

In this work, we will develop a general theoretical framework for these novel types of moiré interfaces and demonstrate, using explicit illustrative examples, the unique potential of this setup.
More specifically, we will derive and study the general conditions under which the electron liquid in the vicinity of the interface can be approximated as crystalline. To this end, we will derive the associated continuum models taking into account the 3D bulk nature of the materials forming the heterostructure. As illustrated schematically in \figref{fig:introfig}(b), we will allow for arbitrary combinations of (in general different) Bravais lattices and lattice parameters meeting at the interface. In addition, we will consider finite twist $\theta$, and take into account that the low-energy degrees of freedom on either side of the interface can be located in the vicinity of different high-symmetry points in the Brillouin zone. To characterize these different location combinations, we introduce the concept of a \textit{crystalline reference point}, defined as a set of parameters $p^*$ characterizing the two Bravais lattices that meet at the interface with the following property: in the limit where the lattice parameters $p$ tend to $p^*$ the electronic low-energy behavior becomes asymptotically crystalline, the usual Bloch theorem can be used, and we obtain electronic bands. For a given distribution of the low-energy electronic degrees of freedom in the respective bulk materials, there are in general multiple continuum reference points, which we therefore characterize by their \textit{momentum}, defined as the required momentum transfer in the interlayer tunneling matrix elements. We identify and discuss all 11, symmetry-inequivalent leading (i.e., of smallest momentum) continuum reference points for the case of one material being cubic (or tetragonal) and the other being hexagonal---the situation relevant for sapphire and SrTiO$_3$ studied in \refcite{MannhartExperiment}. Interestingly, as opposed to the previously studied case of twisting materials with identical lattices, such as twisted bilayer graphene, many of the leading continuum reference points are not at zero twist angle (and zero lattice mismatch) but rather at finite $\theta$. Depending on the reference point, either one or two  quasi-1D channels or mixed dimensionality are realized. The continuum reference points at finite $\theta$ are further associated with minima of the bandwidth (along the direction of the moiré modulation) when varying the twist angle. These magic angles are, however, different from those, e.g., of twisted bilayer graphene \cite{bistritzer2011moire}: they do not crucially depend on the strength of the interlayer tunneling amplitude but are rather of geometric origin, which is why we dub them ``geometric magic angles''.

We further illustrate the physics using explicit model calculations for an interface between a tetragonal and hexagonal lattice. We consider the following combinations of band structures: quadratic bulk bands in both systems, of either particle or hole type, an interface between a topological insulator and bulk materials with a quadratic band, and two topological insulators. Among other interesting band features, such as van Hove singularities and topologically non-trivial bands, we analyze under which conditions isolated, well-localized interface bands appear.

The remainder of the paper is organized as follows. In \secref{GeneralTheory}, we describe the general formalism, which is then applied to the special case of heterostructures built from cubic/tetragonal and hexagonal Bravais lattices, see  \secref{SquareTriangularLattice}. In \secref{ExplicitModelCalcs}, we then present and discuss our explicit band structure calculations, employing different bulk Hamiltonians on either side of the heterostructure. Finally, \secref{ExplicitModelCalcs} summarizes the findings and provides an outlook for future work.

\section{General theory}\label{GeneralTheory}

\subsection{Momentum-space model}\label{MomentumSpaceModel}
We consider two crystalline systems $\ell = 1,2$ and decompose their respective Bravais lattices into a family of lattice planes with distance $d_\ell$; we label the different planes by $Z = n_z d_\ell$, $n_z \in \mathbb{Z}$, and denote the 2D Bravais lattice within each plane by $\text{BL}_\ell$. Let us focus on a finite number $N_b^\ell$ of bands in the energy interval of interest and choose $N_b^\ell$ large enough so as to guarantee the existence of symmetric Wannier states $w_{\ell,\alpha}(\vec{r}-\vec{R},z-Z)$, $\alpha=1,2,\dots N_b^\ell$, $\vec{R} \in \text{BL}_\ell$ for these bands which are exponentially localized around their centers $(\vec{R}+\vec{d}_{\alpha},Z+ z_\alpha)$. Here and in the following, $\vec{r}=(x,y) \in \mathbb{R}^2$ refers to the continuous in-plane coordinates and $z$ to the direction perpendicular to them. 

The heterostructure shown in \figref{fig:introfig}(b) is constructed by stacking the planes with  $Z > 0$ of the material $\ell =2$, twisted by $\theta_2$ along $z$, on top of the $Z \leq 0$ planes of the crystal $\ell =1$, twisted by an angle $\theta_1$ along the same direction. As a result of the twist, the Wannier states with quantum numbers $\ell,\vec{R},Z,\alpha$ are now centered around $(\mathcal{R}_{\theta_\ell}(\vec{R}+\vec{d}_{\alpha}),Z+ z_\alpha)$, where $\mathcal{R}_{\theta_\ell}$ is the orthogonal $2\times 2$ matrix of rotations of 2D vectors by angle $\theta_\ell$. Denoting the associated electronic annihilation operators by $a_{\ell,\vec{R},Z,\alpha}$, the Hamiltonian describing the electronic states in each of the materials is given by
\begin{align}\begin{split}
    \mathcal{H}_{\text{bulk}} &= \sum_{\ell =1 ,2} \sum_{\vec{R},\vec{R}' \in \text{BL}_\ell} \sum_{Z,Z'} \sum_{\alpha,\alpha'=1}^{N_b^\ell} \\ &\qquad a^\dagger_{\ell,\vec{R},Z,\alpha} \hat{h}_{\alpha,\alpha'}^\ell(\vec{R}-\vec{R}',Z,Z') a^\pdagger_{\ell,\vec{R}',Z',\alpha'}, \label{MainTextBulkHamiltonian}
\end{split}\end{align}
where $Z,Z'$ are restricted to their respective values for the given $\ell$ and $\hat{h}_{\alpha,\alpha'}^\ell$ are the tight-binding matrix elements. We will not further specify $\hat{h}_{\alpha,\alpha'}^\ell$ here and refer to \secref{ExplicitModelCalcs} for concrete examples. We note that our formalism also applies to the case where only one of the two materials is 3D, while the other only contains one or few values of $Z$, simply by reducing the sum of $Z$ or $Z'$ in \equref{MainTextBulkHamiltonian} accordingly. This is relevant to 2D materials deposited on 3D bulk substrates, as is relevant, e.g., to the experiments of \refscite{Experiment2Don3D,PhysRevB.107.155418}. 

While translational symmetry along the $Z$ direction is broken due to the interface, \equref{MainTextBulkHamiltonian} is still invariant under in-plane translations by the respective Bravais lattices, $a_{\ell,\vec{R}_0,Z,\alpha} \rightarrow a_{\ell,\vec{R}_0 + \vec{R},Z,\alpha}$ for $\vec{R} \in \text{BL}_\ell$. This motivates performing a partial Fourier transform, $a_{\ell,\vec{R},Z,\alpha} = N_\ell^{-1/2} \sum_{\vec{k}\in \text{BZ}_\ell} e^{i \vec{k} \vec{R}} \bar{c}_{\ell,\vec{k},Z,\alpha}$, where $N_{\ell}$ is the number of 2D unit cells in material $\ell$ (which we assume to be large enough to be able to ignore boundary effects) and $\text{BZ}_\ell$ is the first Brillouin zone associated with $\text{BL}_\ell$. To make momentum conservation more apparent in the interface tunneling terms to be discussed below (see also \appref{DetailedModelDerivation} for more details), we transform back to the unrotated `lab frame', formally achieved by using $c_{\ell,\vec{k},Z,\alpha} := \bar{c}_{\ell,\mathcal{R}_{-\theta_\ell}\vec{k},Z,\alpha}$ instead of $\bar{c}_{\ell,\vec{k},Z,\alpha}$. The Hamiltonian in \equref{MainTextBulkHamiltonian} then assumes the form
\begin{align}\begin{split}
    \mathcal{H}_{\text{bulk}} = \sum_{\ell,Z,Z'} \sum_{\vec{k}\in \text{BZ}_{\ell,\theta_\ell}}  c^\dagger_{\ell,\vec{k},Z} h^\ell(\mathcal{R}_{-\theta_\ell}\vec{k},Z,Z') c^\pdagger_{\ell,\vec{k},Z'}, \label{MomentumSpaceBandHamMainText}
\end{split}\end{align}
which is diagonal in $\vec{k}$. Here we defined $h_{\alpha,\alpha'}^\ell(\vec{k},Z,Z') = \sum_{\vec{R}\in\text{BL}_\ell} e^{i\vec{k}\vec{R}} \hat{h}_{\alpha,\alpha'}^\ell(\vec{R},Z,Z')$, suppressed the $\alpha$ indices, and introduced the rotated Brillouin zone $\text{BZ}_{\ell,\theta_\ell}:=\{\mathcal{R}_{\theta_\ell}\vec{k}\,| \vec{k} \in \text{BZ}_{\ell} \}$.

To describe the coupling between the two different crystals across the interface, we assume that only the degrees of freedom in the layers closest to the interface, i.e., $Z=0$ ($Z=d_2$) in material $\ell = 1$ ($\ell = 2$) in \figref{fig:introfig}(b) couple significantly. Apart from the dependence on the type of Wannier states ($\alpha_{1,2}$ in the equation below), the tunneling amplitudes $t$ should only dependent on the difference between the Wannier centers. The corresponding Hamiltonian then reads as
\begin{align}\begin{split}
    \mathcal{H}_{\text{inter}} &= \sum_{\vec{R}_\ell \in \text{BL}_\ell} \sum_{\alpha_\ell=1}^{N_b^\ell} a^\dagger_{1,\vec{R}_1,Z=0,\alpha_1} a^\pdagger_{2,\vec{R}_2,Z=d_2,\alpha_2} \\ & \quad \times\frac{1}{\sqrt{N_1 N_2}}t_{\alpha_1,\alpha_2}(\mathcal{R}_{\theta_1}\vec{R}_1 - \mathcal{R}_{\theta_2}\vec{R}_2)   + \text{H.c.}. \label{RealSpaceTunnelingHam}
\end{split}\end{align}
Note that $N^1_b \neq N^2_b$ is possible, since the two materials on either side of the interface can be different, such that $t$ (and also $T$ below) will not be square.
Performing, again, a partial Fourier transform and introducing the Fourier series expansion of the tunneling matrix elements, $t_{\alpha_1,\alpha_2}(\vec{R}) =  \sum_{\vec{q}} (T_{\vec{q}})_{\alpha_1,\alpha_2} e^{i\vec{q}\vec{R}}$, the interface part (\ref{RealSpaceTunnelingHam}) of the Hamiltonian becomes
\begin{align}\begin{split}
    \mathcal{H}_{\text{inter}} &= \sum_{\vec{k}_{\ell}\in\text{BZ}_{\ell,\theta_\ell} } \sum_{\vec{G}_\ell \in \text{RL}_{\ell,\theta_\ell}} \sum_{\alpha_\ell=1}^{N_b^\ell} \delta_{\vec{k}_1 +\vec{G}_1,\vec{k}_2 +\vec{G}_2} \\ & c^\dagger_{1,\vec{k}_1,Z=0,\alpha_1} (T_{\vec{k}_1 +\vec{G}_1})_{\alpha_1,\alpha_2} c^\pdagger_{2,\vec{k}_2,Z=d_2,\alpha_2}  + \text{H.c.}. \label{MomentumSpaceTunnelingMainText}
\end{split}\end{align}
Here, $\text{RL}_{\ell,\theta_\ell}$ is the rotated reciprocal lattice, i.e., $\text{RL}_{\ell,\theta_\ell} := \{\mathcal{R}_{\theta_\ell}\vec{G}| \vec{G}\in \text{RL}_\ell\}$, where $\text{RL}_\ell$ is the reciprocal lattice of $\text{BL}_\ell$. From \equref{MomentumSpaceTunnelingMainText} we can see that, as expected, the tunneling term only conserves momentum modulo reciprocal lattice vectors. However, since we consider two in general different materials with different Bravais and, hence, reciprocal lattices in the lattice planes, the interlayer tunneling leads to scattering between different momenta in the respective Brillouin zones---even without any twist $\theta_\ell = 0$. Therefore, for generic parameters, the effective electronic Hamiltonian cannot be approximated as crystalline in the vicinity of the interface. Nonetheless, as we show next, there are specific values of the relative twist angle and lattice mismatch where the system does become asymptotically crystalline; as opposed to the commonly studied case of twisting the same two materials, these specific reference points in parameter space, which crucially depend on the localization of the low-energy electronic states in the Brillouin zones, are not necessarily at vanishing twist angle and lattice mismatch.

\subsection{Crystalline reference points}
Let us assume we are interested in the behavior of the system in a small energy window (e.g., around the Fermi level) such that only a small fraction of the values of $\vec{k}\in \text{BZ}_{\ell,\theta_\ell}$ in \equref{MomentumSpaceBandHamMainText} need to be taken into account. Then we can replace $\mathcal{H}_{\text{bulk}}$ in \equref{MomentumSpaceBandHamMainText} by the patch model
\begin{equation}
    \mathcal{H}_{\text{bulk}}^{\text{P}} = \sum_{\ell,Z,Z'} \sum_j \sum_{\vec{q}, |\vec{q}| < \Lambda}  f^\dagger_{\ell,\vec{q},Z,j} h_j^\ell(\mathcal{R}_{-\theta_\ell}\vec{q},Z,Z') f^\pdagger_{\ell,\vec{q},Z',j}. \label{PatchHamiltonian}
\end{equation}
Here $j$ labels the different high-symmetry points $\vec{P}_{j,\ell}$ in the respective rotated Brillouin zone $\text{BZ}_{\ell,\theta_\ell}$ around which $h^\ell(\mathcal{R}_{-\theta_\ell}\vec{k},Z,Z')$ has eigenvalues in the energy range of interest. We further introduced the patch fermions $f_{\ell,\vec{q},Z,j} := c_{\ell,\vec{P}_{j,\ell}+\vec{q},Z}$ and Hamiltonian $h_j^\ell(\vec{q},Z,Z') := h^\ell(\mathcal{R}_{-\theta_\ell}\vec{P}_{j,\ell}+\vec{q},Z,Z')$. 

For generic values of the twist angle $\theta = \theta_1 - \theta_2$, ratios $\eta_{n} := |\vec{b}_{n,2}|/|\vec{b}_{n,1}|$, $n=1,2$, of the lattice constants ($\vec{b}_{n,\ell}$ are the basis vectors of $\text{RL}_{\ell,\theta_\ell}$), and possibly angles $\phi_\ell$ for monoclinic lattices, one needs large $\vec{G}_\ell$ (much longer than $|\vec{b}_{n,\ell}|$) to scatter between the relevant electronic degrees of freedom that are kept in \equref{PatchHamiltonian} at finite but small values of the momentum cutoff $\Lambda$; more mathematically, one has to pick specific and, for $\Lambda \rightarrow 0$, large $\vec{G}_\ell$ such that the magnitude of the momentum transfer in the patch theory,
\begin{equation}
    \delta b := |\delta \vec{b}|, \quad \delta \vec{b} = \vec{P}_{j_2,2} + \vec{G}_2 -(\vec{P}_{j_1,1}+\vec{G}_1), \label{DefinitionOfdeltab}
\end{equation}
happens to become sufficiently small for some pair $(j_1,j_2)$ of high symmetry points. Since we expect $T_{\vec{q}}$ in \equref{MomentumSpaceTunnelingMainText} to decay with increasing $|\vec{q}|$, the mutual impact of the two lattices at the interface is weak for small $\Lambda$. What is more, one also typically expects that there are multiple $\vec{G}_\ell$ of similar magnitude where $\delta b$ is sufficiently small, leading to multiple incommensurate momentum transfers $\delta \vec{b}$ in the patch theory. As a consequence, the interface system cannot be approximated as a crystal (see \appref{AwayFromAContinuumReferencePoint} for a more detailed demonstration in an example). 

To study well-defined crystalline limits, we define \textit{crystalline reference points} as values $p^*=(\eta^*_n,\theta^*,\phi^*_\ell)$ of the lattice mismatch, twist angle, and possibly angle(s) for monoclinic lattices where there are $\vec{G}_\ell \in \text{RL}_{\ell,\theta_\ell}$ of finite length with $\delta b = 0$ in \equref{DefinitionOfdeltab} for at least one pair $(j_1,j_2)$. Moreover, to quantify the impact of the moiré modulation, we consider 
\begin{equation}
    Q(\vec{G}_\ell,j_\ell) := \text{max}_{\ell=1,2} |\vec{P}_{j_\ell,\ell} + \vec{G}_\ell|, \label{DefinitionOfQGen}
\end{equation}
which, right at the crystalline reference point and for $\Lambda \rightarrow 0$, coincides with the magnitude of the argument of $T_{\vec{k}_1+\vec{G}_1}$ of the corresponding term in \equref{MomentumSpaceTunnelingMainText}. We call the minimal value of $Q(\vec{G}_\ell,j_\ell)$ for any $\vec{G}_\ell$, $(j_1,j_2)$ for which $\delta b =0$ in \equref{DefinitionOfdeltab} \textit{the wavevector} $Q$ of the reference point.  

A crystalline reference point has physical significance, since, in the limit $\Lambda \rightarrow 0$, only configurations with $\delta b \rightarrow 0$ contribute. Due to the aforementioned decay of $T_{\vec{q}}$ with $|\vec{q}|$, we restrict the sum over $\vec{G}_\ell$ and $(j_1,j_2)$ in \equref{MomentumSpaceTunnelingMainText} to those configurations with smallest $Q(\vec{G}_\ell,j_\ell)$. As a result of the discrete nature of the reciprocal lattices, this set of configurations is still well-defined and unchanged when detuning the lattice parameters $(\eta_n,\theta,\phi_\ell)$ slightly away from the crystalline reference point and using a small but finite cutoff $\Lambda$. The main difference now is that the momentum transfer $\delta \vec{b}$ in the patch theory, given in \equref{DefinitionOfdeltab}, becomes finite as well; this defines the emergent moiré superlattice modulation that the electrons experience in the vicinity of the interface.
While point symmetries can lead to degeneracies in $Q(\vec{G}_\ell,j_\ell)$ such that multiple $\vec{G}_\ell$ and $(j_1,j_2)$ have to be included, we start from crystalline materials such that only crystalline point symmetries can emerge as exact symmetries at the interface. This is why the resulting set of $\delta \vec{b}$ that are to be included as discussed above lead to a moiré superlattice with crystalline translational symmetry;
we have also checked this explicitly in all examples studied in this paper. 

\begin{figure}[tb]
   \centering
    \includegraphics[width=1.0\linewidth]{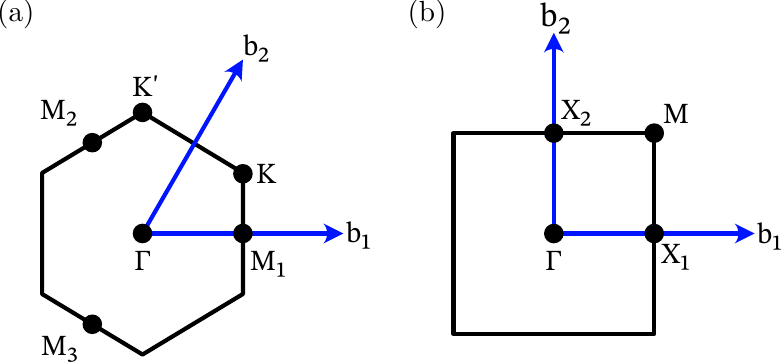}
    \caption{(a) and (b) show the basis vectors, the first Brillouin zone and the high-symmetry points of the triangular and square lattices, respectively.}
    \label{fig:latticesgeneral}
\end{figure} 

Before we move on to the main focus of this work, the case where the systems on either side are different (and 3D), let us illustrate these general considerations using one of the most well-known examples of a moiré superlattice system---twisted bilayer graphene \cite{dos2007graphene,MeleModel,bistritzer2011moire,dos2012continuum}. In that case, each of the two materials $\ell = 1,2$ is just a single layer, formally taken into account by ignoring the arguments $Z$, $Z'$ in \equsref{MomentumSpaceTunnelingMainText}{PatchHamiltonian}. In graphene, the degrees of freedom close to the Fermi level are well-localized around the K and K' points of the respective Brillouin zone of the underlying triangular lattice, see \figref{fig:latticesgeneral}(a), i.e., $\vec{P}_{1,\ell} = \mathcal{R}_{\theta_\ell} \text{K}$ and $\vec{P}_{2,\ell} = \mathcal{R}_{\theta_\ell} \text{K'}$. In this case, vanishing twist angle and identical lattice constants define a crystalline reference point simply because it implies $\vec{P}_{j,1} = \vec{P}_{j,2}$ and, thus, $\delta b =0$ as long as $\vec{G}_1=\vec{G}_2$ and $j_1 = j_2$ in \equref{DefinitionOfdeltab}. There are exactly three values $(\vec{G}_1,\vec{G}_2) = (0,0)$, $(\vec{b}_1,\vec{b}_1)$ and $(\vec{b}_2,\vec{b}_2)$ where $Q(\vec{G}_\ell,j_\ell)$ in \equref{DefinitionOfQGen} is minimal and given by $|\text{K}|=|\text{K'}|$, which is thus the wavevector of this continuum reference point. For finite, but small twist angles $\theta$ (and assuming $\theta_1=-\theta_2 = \theta/2$ for notational simplicity), we obtain three non-zero values of $\delta \vec{b}$ which, for $j_1=j_2=1$, read as $\delta \vec{b} \sim |\vec{b}_1|(\theta/\sqrt{3},-\theta)/2^T$, $\delta \vec{b} \sim |\vec{b}_1|(\theta/\sqrt{3},\theta)/2^T$, and $\delta \vec{b} \sim |\vec{b}_1|(-\theta/\sqrt{3},0)^T$ up to linear order in $\theta$. These are just related by three-fold rotation such that the emergent moiré lattice is also a triangular lattice. We will next discuss in detail that two non-identical lattices can lead to non-trivial crystalline reference points located at finite twist angles and/or lattice mismatch.

\section{Square and triangular lattice}\label{SquareTriangularLattice}
For concreteness, we next focus on the case in which $\text{BL}_{1}$ is a square and $\text{BL}_{2}$ a triangular lattice. 
This choice is not only motivated by its relevance to recent experiments \cite{MannhartExperiment} but also because they represent the simplest non-trivial example of two distinct Bravais lattices, which are only parametrized by two relative parameters, the twist angle $\theta$ and lattice mismatch $\eta=\eta_1=\eta_2$ \footnote{Here we drop the subscript index since all the basis vectors from both layers have the same size.}. 

First, let us look in detail at an illustrative example. We consider the situation in which the low energy modes are localized around the around the $\Gamma$ point in the triangular layer, \figref{fig:latticesgeneral}(a), and around the $X$ points in the square lattice, \figref{fig:latticesgeneral}(b). Subsequently, we will generalize the following reasoning for all the possible combinations of high-symmetry points and summarize these results in \tableref{table:SquareAndTriangularLattice}.

In our example, we note that the square lattice should have low energy modes around both the $X_1$ and $X_2$ points due to (and related by) $C_{4}$ symmetry. However, we cannot consider just one of these points and argue that the other one should be equivalent (up to a $90^{\circ}$ rotation transformation) in the moir\'e interface. The reason for this inequivalence is the coupling to the second layer, which has a triangular lattice and does not have $C_4$ symmetry. Therefore, we need to look at $X_1$ and $X_2$ separately.

Now, with this detail in mind, we search for the reference point from which small deviations will lead to the strongest moiré modulation. To this end, we search for $\vec{G}_1$ and $\vec{G}_2$ that will make $\delta b\rightarrow 0$ with the lowest $Q$ in \equref{DefinitionOfQGen} possible. As we demonstrate more explicitly in \appref{ap:more_shells}, this can be analyzed by using a finite number of shells of $\vec{G}_{1,2}$ around the origin: once a continuum reference point is found with a sufficiently small $Q$, adding more shells can only lead to larger $Q$ and, hence, subleading reference points. 
To keep the discussion compact, we restrict ourselves to small twist angles $|\theta| \leq 5^{\circ}$, while larger twist can be straightforwardly included. We note that much larger $|\theta|$ would just lead to a relabeling of high symmetry points (e.g., $\theta \sim 90^\circ \Rightarrow X_1 \leftrightarrow X_2$). We also only allow (reciprocal) lattice mismatch $\leq 20\%$, i.e., $ 0.8 \leq  \eta \leq 1.2$.

\begin{figure}
    \centering
    \includegraphics[width=\linewidth]{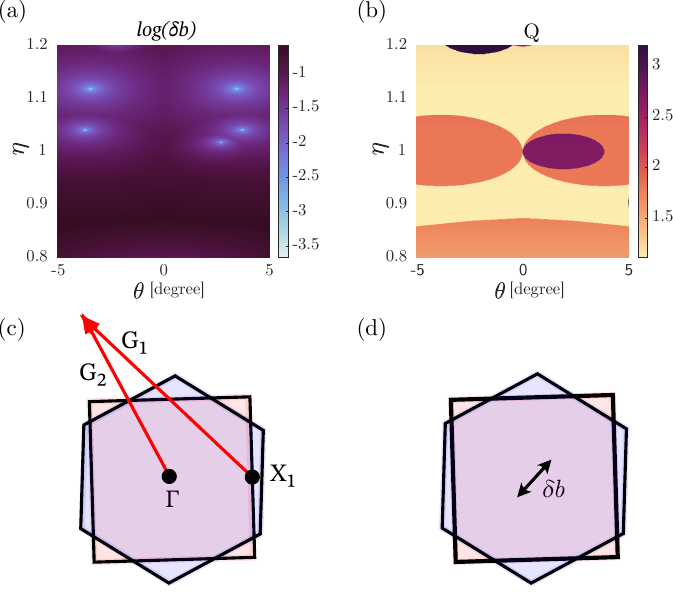}
    \caption{(a) Minimal $\delta b$ as a function of twist angle $\theta$ and lattice mismatch $\eta$. Here we consider the high-symmetry point $X_1$ on the square lattice and $\Gamma$ on the triangular lattice as expansion points. For a set of isolated points, $\delta b \rightarrow 0$, defining the continuum reference points. In (b) we show the respective $Q$, which identifies the leading reference points at $\theta = \pm 3.43^{\circ}$ and $\eta = 1.118$ with momentum $Q=1.12$. For clarity, we considered only two shells of lattice points for this specific figure. (c) Brillouin zones at the leading crystalline reference point for $X_1, \Gamma$. Here, $G_1$ and $G_2$ are the smallest reciprocal lattice vectors (and so are $-G_{1,2})$ that fulfill the condition $G_1 + X_1 = G_2 + \Gamma$. (d) The vectors $\delta b$ for a small displacement from the reference point. We can see that they are related through $C_{2z}$ symmetry.}
    \label{fig:q_scan}
\end{figure}

This search is done through a scan in the range of twist angle and lattice mismatch. For each pair $(\theta, \eta)$, we find $\vec{G}_1, \vec{G}_2$ that give us the smallest $\delta b$, which we plot in \figref{fig:q_scan}(a). We look at the set of $(\theta', \eta')$ that makes $\delta b \rightarrow 0$ (i.e., $X_{1,2} + \vec{G}_1' = \Gamma + \vec{G}_2'$ for some $\vec{G}_1', \vec{G}_2'$ )  and compare their $Q$ in \figref{fig:q_scan}(b). Among this set, we take the pair of distortion parameters $(\theta^*,\eta^*)$ with the smallest $Q$ and consider it to be the crystalline reference point for that given combination of high-symmetry points.

\begin{table*}[tb]
\begin{center}
\caption{We here consider the different combinations of high-symmetry points of the square and triangular lattice (first two columns), around which the low-energy electronic degrees of freedom are localized in the respective bulk materials. The third and fourth columns show the respective leading crystalline reference points, i.e., the twist angle and lattice mismatch necessary to make $\delta\textbf{b}\rightarrow 0$ with the lowest momentum $Q$ possible; the associated value of $Q$ is shown in the fifth column. This value is the magnitude of the argument of the interlayer coupling $T_{\vec{q}}$ in \equref{MomentumSpaceTunnelingMainText} and is a natural measure for how strong the moir\'e modulation is (lower $Q$ generally means a larger coupling). In the last column, we summarize the main features of the resulting moir\'e band structure for each combination, where ``rec.'' stands for reconstruction; see main text for details. We considered $|\theta|\leq 5^\circ $ and $0.8 \leq  b_t/b_s  \leq 1.2$.}
\label{table:SquareAndTriangularLattice}
\begin{ruledtabular}
 \begin{tabular}{cccccc} 
\multicolumn{2}{c}{high-symmetry point} & \multicolumn{2}{c}{crystalline reference point} & \multicolumn{2}{c}{Properties}
\\ \cline{1-2} \cline{3-4} \cline{5-6} 
Square & Triangular & $\theta^*$ & $\eta^*=b_t/b_s$ &  $Q/b_s$ & Band structure
\\ \hline
$\Gamma$ & $\Gamma$ & $0^\circ$ & $1$  & 0, 1 & 1D with $C_{2z}$ \\
$\Gamma$ & $M$ & $0^\circ$ & $2/\sqrt{3}$ & 1 & $\Gamma M_1$:1D with $C_{2z}$; $\Gamma M_{2,3}$: no rec. \\
$\Gamma$ & $K$ & $0^\circ$ & $\sqrt{3}/2$ & 1 & 1D w/o $C_{2z}$, $\Gamma K$ $\xleftrightarrow{C_{2z}}$ $\Gamma K'$   \\ \hline
$M$ & $\Gamma$ & $\pm 4.1^\circ$ & $0.801$ & 2.12 & 1D with $C_{2z}$ \\
$M$ & $M_1$ & $\pm 2.36^\circ$ & $0.877$  & 1.58 & $M M_1$: 1D with $C_{2z}$; $M M_{2,3}$: no rec.  \\
 & $M_{3(2)}$ & $ (-)4.54^\circ$ & $0.877$ & 1.58 & $M M_{3(2)}$: 1D with $C_{2z}$; $M M_{2(3)}$: no rec.\\
 & $M_{3(2)}$ & $ (-)0.67^\circ$ & $1.195$ & 1.58 & $M M_{3(2)}$: 1D with $C_{2z}$; $M M_{2(3)}$: no rec.  \\
$M$ & $K$ & $\pm0.66^\circ$ & $1.035$   & 1.58 & 1D w/o $C_{2z}$, $M K$ $\xleftrightarrow{C_{2z}}$ $M K'$ \\ \hline
$X$ & $\Gamma$ & $\pm3.43^\circ$ & $1.118$  & 1.12 & $X_1\Gamma$: 1D ($C_{2z}$); $X_2\Gamma$: no rec. \\
$X$ & $M$ & $0^\circ$ & $1$ & 0.5 & $X_1 M_1$:1D with $C_{2z}$; other comb.: no rec. \\
$X$ & $K$ & $0^\circ$ & $\sqrt{3}/2$  & 0.5 &$X_2 K^{(\prime)}$: 1D w/o $C_{2z}$, $X_2 K$ $\xleftrightarrow{C_{2z}}$ $X_2 K'$; $X_1 K^{(\prime)}$: no rec. \\
\end{tabular}
\end{ruledtabular}
\end{center}
\end{table*}

By performing this scan for the first combination $\vec{P}_{1,1}=X_1$, $\vec{P}_2=\Gamma$, we obtain several crystalline reference points with different orders ($Q$), see minima in \figref{fig:q_scan}(a). We find the one with leading order at a twist angle of $\theta^* =\pm 3.43^\circ$ (the two signs are just related by mirror symmetry) and scaling $\eta^* = 1.118$. Here, $Q=1.12 b_s$, where $b_s = |\vec{b}_{1,i}|$ is the length of primitive vectors of the reciprocal lattice of the square-lattice material $\ell =1$. We can visualize the condition $\delta b \rightarrow 0$ in  \figref{fig:q_scan}(c) through  $\vec{G}_1 + X_1$ and $\vec{G}_2 + \Gamma$. There, we can see that these points are the same. The explicit form of the lattice points is $\vec{G}_1=(-1,1)b_s$ and $\vec{G}_2=(-\frac{1}{2}, \frac{\sqrt{3}}{2}) b_t$, $b_t = \eta^*  b_s$, which are afterwards rotated by $\theta^*$. The choice $-\vec{G}_1$ and $-\vec{G}_2$ also fulfill the wanted condition.

Note that the vectors $\pm(\vec{G}_1, \vec{G}_2)$ are collinear and related by $C_{2z}$. Hence, for a small deviation, we find two $\delta \textbf{b}$ with the same properties as this set of lattice vectors, as we can see in \figref{fig:q_scan}(d). We obtain an explicit form for $\delta \vec{b}$ by performing a linear expansion around $(\theta^*,\eta^*)$, $\delta\vec{b}\approx (-0.47 \epsilon_\eta -0.98 \epsilon_\theta , -0.88 \epsilon_\theta + 0.53 \epsilon_\eta )$, where $\epsilon_{\theta}=\theta -\theta^*$ and $\epsilon_{\eta}=\eta -\eta^*$ represent small deviations from the continuum reference point. 
Based on these observations, we obtain a band that is reconstructed by the moir\'e lattice only in one direction and has $C_{2z}$ symmetry [denoted by 1D($C_{2z}$) in \tableref{table:SquareAndTriangularLattice}]. We further note that in \figref{fig:q_scan}(a-b) the $C_{2z}$ symmetry is apparently not present. The seeming asymmetry is a consequence of using a finite number of shells (since we are interested in the leading order crystalline reference point) and would vanish if infinitely many shells were considered.

For the second combination $\vec{P}_{1,2}=X_2$, $\vec{P}_2=\Gamma$, we repeat the same steps but this time we find the leading crystalline reference point at $\theta^*=0^{\circ}$ and $\eta^* = \sqrt{3}/2$ with $Q=1.5$. Therefore, by noting that $X_1,\Gamma$ and $X_2,\Gamma$ have different reference points  \footnote{Even considering more shells, the point $\theta^* = 3.43^\circ, \eta^* = 1.118$ does not become a reference point for $X_2,\Gamma$.}, we conclude that only one of these combinations will lead to a sizable moiré modulation (for small deviations from one of the reference points). In addition, the moir\'e modulation has generically a stronger effect for $X_1,\Gamma$ since it has the smaller momentum $Q$; this is why we include $\theta^* =\pm 3.43^\circ$, $\eta^* = 1.118$ as the leading order crystalline reference point in \tableref{table:SquareAndTriangularLattice}.

Now, repeating this procedure for the remaining combinations of high-symmetry points we obtain Table \ref{table:SquareAndTriangularLattice}. We can see all the leading crystalline reference points, the respective $Q$ values and what type of band reconstruction we obtain. Let us discuss remarkable features of the band structures found in Table \ref{table:SquareAndTriangularLattice}. First, we notice that all the combinations involving the $M$ points from the triangular lattice lead to only one of the valleys being reconstructed. All these reconstructed bands are quasi-1D and have $C_{2z}$ symmetry. For the  combination $(M,M_{2,3})$, we found two degenerate reference points and therefore included both.
Furthermore, also when the $X$ point (of the square lattice) is involved, just a subset of pockets is reconstructed. These cases are qualitatively different from other moiré platforms, in which one can obtain either quasi-1D (e.g., twisted $BC_3$ \cite{PhysRevB.107.085127}) or 2D modulation (e.g., twisted bilayer graphene \cite{dos2007graphene}). Here, we observe ``mixed dimensionality'' arising.

Secondly, we see that all the combinations involving $K(K')$ lead to two 1D channels, which, akin to the two valleys in twisted bilayer graphene \cite{dos2007graphene,MeleModel,bistritzer2011moire,dos2012continuum}, do not have $C_{2z}$ symmetry individually but are related by it. This phenomenon is also similar to the case of twisted $BC_3$ \cite{PhysRevB.107.085127}, which has low-energy modes around the $M$ points. There all the valleys exhibit 1D modulation and are related through $C_{3z}$.

Thirdly, we note some results with simple geometrical interpretations. In the case of $(\Gamma, \Gamma)$, we see that $Q=0$ is possible, if one chooses $\vec{G}_{1,2}=0$; however this choice implies $\delta b = 0 $ for all twist angles and lattice mismatches while not being associated with a moir\'e modulation. For this reason, one should also include the combination of $\vec{G}_{1,2}$ with next-smallest $\delta b$ and associated value of $Q=b_s$ in \equref{DefinitionOfQGen}, which we also list in \tableref{table:SquareAndTriangularLattice}. 
%Hence it does not lead to any moiré modulation. 
Other scenarios with simple 
visualization are the ones with $\theta = 0^\circ$. Here, $\eta$ is enough to make $\vec{P}_{j_2,2}=\vec{P}_{j_1,1}$. If $\eta=\sqrt{3}/2$ (for $\Gamma$-$K$ or $X$-$K$), the Brillouin zone of the triangular lattice is inscribed in the square lattice Brillouin zone. On the other side, if $\eta = 2/\sqrt{3}$ (relevant to $\Gamma$-$M$), the two real spaces lattices have basis vectors with the same norm.

We remark that for all cases of the square/triangular lattice interface continuum reference points appear as \textit{isolated} points. There are no lines of reference points because we only deal with two parameters ($\eta, \theta$) instead of three. Therefore, choosing two vectors $\vec{G}_1$ $\vec{G}_2$ will define a point.
On the other hand, the square/rectangular lattice interface, for instance, would be parameterized by $\eta_1, \eta_2$ and $\theta$, since the rectangular lattice basis vector have different length. Let us look at an example for this interface to see how a \textit{line} of crystalline reference points can emerge. Consider both layers with modes around the respective $\Gamma$ points. We obtain one trivial reference point for 2D modulation at $\eta_1=\eta_2=1$ and $\theta=0$, i.e., when the rectangular lattice becomes a square lattice. However, it is easy to see that there is a line of reference points, all with quasi-1D moiré modulation, whenever either $\eta_1$ or $\eta_2$ is 1, and $\theta=0$.

\section{Explicit model calculations}\label{ExplicitModelCalcs}
In the following, we apply the general framework to concrete models. We build on the geometric arguments of \secref{SquareTriangularLattice} and investigate the quasi-1D electronic bands arising from the moir\'e modulations at the interfaces of 2D as well as 3D materials with square and triangular lattice symmetries. At the end, we will discuss how the presence of non-trivial crystalline reference points leads to a novel type of ``magic angle'' with flat bands.

\subsection{NI-NI at $(\Gamma, \Gamma)$}\label{sec:Expl_calc_NINI}
We start the discussion with the simple case of non-topological 3D band-insulators  in each layer, referred to as normal insulators (NI) in the following. We model them with quadratic dispersions given by $\epsilon_{\vec{k}}^{\ell} =\vec{k}^2/2m^*_\ell - \mu_\ell$. We always restrict the calculations to a single relevant band in each layer and assume pockets around the $\Gamma$ points such that the low-energy physics is dominated by momenta in the vicinity of $\vert \vec{{k}}\vert<\Lambda$. This case is described by the first row in \tableref{table:SquareAndTriangularLattice}. The $Q=0$ contribution corresponds to an interaction of unfolded bands, \ie $\vdb=0$ and gives rise to the coupling term $T_{\vec{k}}$ in \equref{MomentumSpaceTunnelingMainText}, which we approximate as a constant $T_{\vec{k}}=T_0$. The $Q=1$ parts yield a finite moiré modulation with wave vectors $\pm\vdb$ and give rise to couplings between momenta which are separated by $\pm\vdb$. We approximate the associated amplitudes by the constant parameter $T_1$. Additionally accounting for the 3D structure of the individual materials by adding nearest-neighbor hopping along the $z$ direction, the Hamiltonian can be written as
\begin{align}\label{eq:hamiltonian_nini}
    \mathcal{H}_\text{tNi} =& \sum_{|\vec{k}| < \Lambda}\sum_{Z=0}^{N_Z-1} c^\dagger_{\ell,\vec{k},Z} \begin{pmatrix} \epsilon_{R_{-\theta_1}\vec{k},1} & T_{0}\delta_{Z,0} \\ T^*_{0}\delta_{Z,0} & \epsilon_{R_{-\theta_2}\vec{k},2} \end{pmatrix}_{\ell,\ell'}  c^\pdagger_{\ell',\vec{k},Z}\\ \nonumber
    &+ \left[ T_1 \sum_{|\vec{k}| < \Lambda}\sum_{\pm} c^\dagger_{1,\vec{k},Z=0} c^\pdagger_{2,\vec{k} \pm \delta\vec{b},Z=0} + \text{H.c.} \right]\\
    \nonumber
    &+\left[ t_z \sum_{|\vec{k}| < \Lambda}\sum_{\ell}\sum_{Z=1}^{N_Z-1} c^\dagger_{\ell,\vec{k},Z+1} c^\pdagger_{\ell,\vec{k},Z} + \text{H.c.} \right]
\end{align}
where $t_z$ is the amplitude of the vertical hopping. For notational convenience, we here use $Z = 0,1,2,\dots$ as the number of lattice planes between the associated states and the interface layer; as such the two planes closest to the interface are at $Z=0$. 
%The interface is located at $Z=0$. For notational convince we the sum over the discrete vertical indices $Z$ only covers positive values. 
Nevertheless we model the system such the lower layer $\ell=1$ is located at negative vertical coordinates $z<0$, as in \figref{fig:introfig}(b).

In the numerical computations, the heterostructure is modeled with $N_Z/2=14$ layers of each material. We include $10$ moir\'e Brillouin zones ($N_b=21$ $\vdb$-vectors), centered around zero so as to respect all symmetries. The coordinates are chosen such that $\vdb = (\db, 0)^T$ points along the $k_x$ direction. We set $\db=1$ (as our momentum scale) and define the energy scale $\db^2/2m_1^*$ in which all energies are measured in this subsection. The chemical potentials and coupling terms are given by $\mu_1=0$, $\mu_2=-1.8$ and $T_0=1$, $T_1=0.8$, respectively. The effective mass in the $\ell=2$ layer is given by $|m_2^*/m_1^*|=1.3$.

\begin{figure}[tb]
    \centering
    \includegraphics[width=\linewidth]{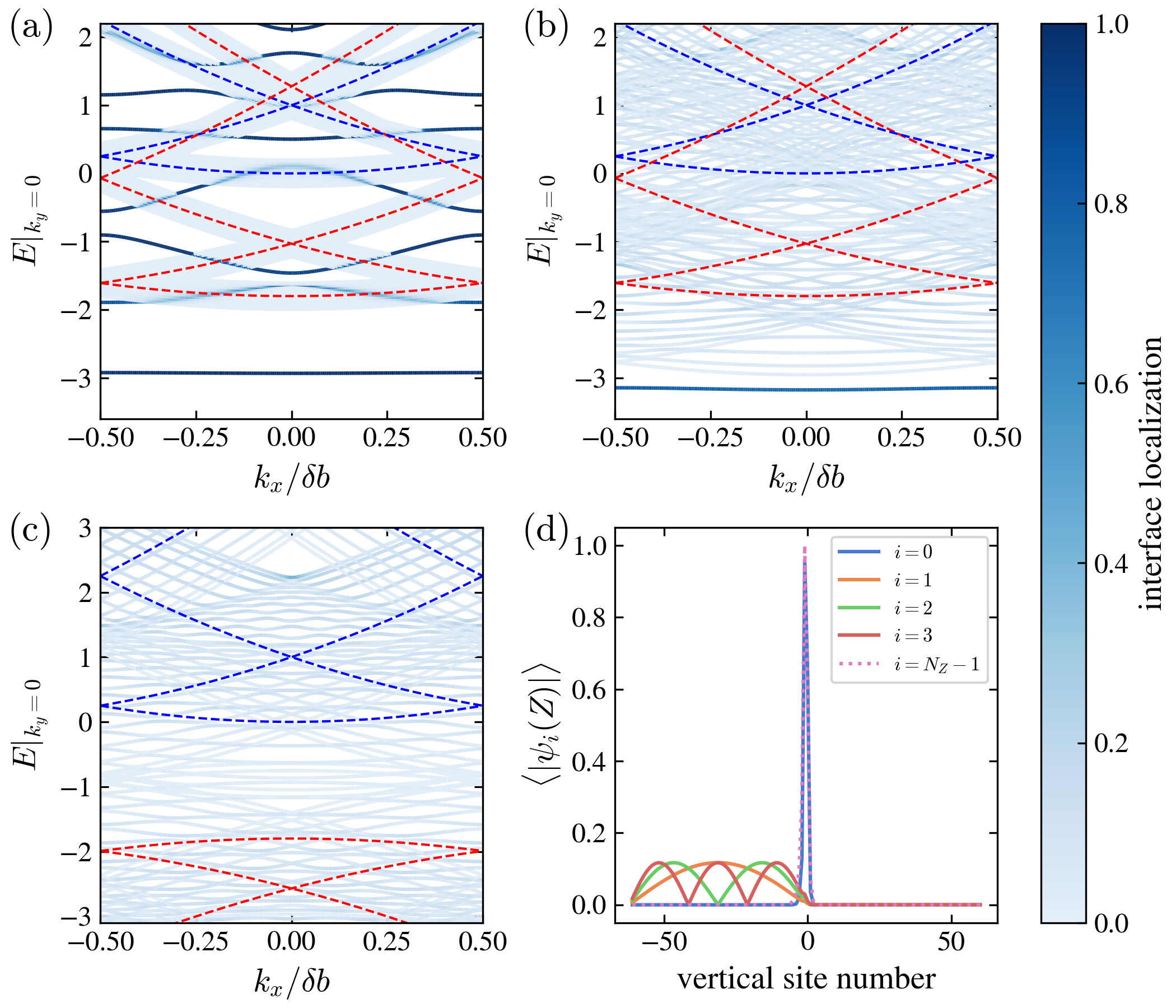}
    \caption{Band structure of twisted NIs, as described by \equref{eq:hamiltonian_nini}, for small (a) and order-one (b) vertical hoppings $t_z=0.1$ and $t_z=0.6$ compared to the interface interaction terms $T_0=1$ and $T_1=0.8$ for two particle-like bands. Panel (c) shows the same band structure as panel (b), except for bands in the lower layer being hole-like. The dashed lines refer to the unreconstructed bands at $t_z=0$. The vertical localization of the  wave functions corresponding to $\vec{k}=(-0.4/\delta b,0)^T$ in (a) is shown in panel (d).}
    \label{fig:NiNi}
\end{figure}

We can see the resulting band structures for particle-like bands ($m_\ell^*>0$) in \figref{fig:NiNi}(a) and (b). There the bare dispersion of the individual layers is represented by dashed lines. The solid lines are obtained by diagonalizing the Hamiltonian \eqref{eq:hamiltonian_nini} including the interface terms $T_0$ and $T_1$. The color represents the interface localization of the respective wave functions. A value of one means that a state is entirely located at the interface. Panel (a) shows the limit of a layered material where the vertical intralayer hopping $t_z=0.1$ is significantly smaller than the interlayer hopping terms at the interface. In this limit we find two distinct types of bands, \ie interface-localized bands as well as 3D bulk bands. While the former originate from avoided crossings caused by the $T_{0,1}$ terms, the latter are not localized and arise from lifting the degeneracy of the bare bands due to the vertical hopping $t_z$. Figure~\ref{fig:NiNi}(d) represents the localization of the first few wave functions by plotting their layer-averaged absolute values as a function of vertical site number $z/d_\ell$ at $\vec{k}=(-0.4/\db, 0)^T$. The remaining parameters are equivalent to the ones used for panel (a) such that the plots are directly comparable. Note that the energetically lowest band $(i=0)$ has a sharp peak at $Z=0$ while the wave functions of the subsequent bands form standing waves along the vertical axis of the twisted system. All shown states appear in the lower layer \ie $\ell=1 \Leftrightarrow z\leq0$; this asymmetry primarily results from the asymmetry in the chosen chemical potential values in the two materials. %as a consequence of the choice of the chemical potential in this layer. 
Each (bare) band gives rise to $N_Z/2-1$ bulk bands (each crystal consists of $N_Z/2$ vertical layers, one of them is at the interface). For the specific value of $\vec{k}$ in \figref{fig:NiNi}(d) there are two sets of $N_Z/2-1$ bulk bands after which another localized interface band appears.

Figure~\ref{fig:NiNi}(b) shows the band structure in the more generic situation of $t_z=0.6$ being of the same order as $T_{0,1}$. The bulk bands hybridize with almost all localized bands. However, the lowest band is a persistent interface mode. Due to the particle-particle nature of the bands the spectrum is bounded from below. Hence the lowest energy band is pushed down and avoids crossings with the bulk bands emerge in a rather broad parameter regime. This interface mode can merge with the bulk modes and disappear once the vertical hopping dominates the energy scale. Figure~\ref{fig:NiNi}(c) shows the same situation as \figref{fig:NiNi}(a) with a hole-like band in the upper layer \ie $m^*_{\ell=2}<0$. As the spectrum is now unbounded in both directions the interface localized band is absorbed by the bulk and the vertical hopping term dominates the band structure even for smaller values of $t_z$.

\subsection{TI-TI at $(\Gamma, \Gamma)$}\label{sec:Expl_calc_TiTi}
The next model we investigate consists of two twisted TIs. As opposed to previous works \cite{PhysRevB.106.075142,KoshinoTwistedTIs}, we here discuss heterostructures formed from different materials with different Bravais lattices. As above, we will  focus on square and triangular lattices and assume, for concreteness, that the surface  Dirac cones are centered around the $\Gamma$ point in both materials. 

As described more formally in \appref{sec:app_projection_TI_surface}, once the bulk materials are projected onto their respective surface states, one obtains an effectively 2D model.
Another crucial difference compared to the discussion of 3D bulk NIs in the previous subsection is that now $c_{\ell,\vec{k}}$ are two-component operators in pseudospin space with Pauli matrices $\sigma_j$; this means that, in both layers, $\sigma_j$ transform in the same way as spin operators. The bare dispersions in \equref{eq:hamiltonian_nini} are thus promoted to $2\times2$ operators $\epsilon_{\vec{k}}^\ell\rightarrow h^{\ell}_{\vec{k}} = v_\ell \left[ \sigma_y k_x - \sigma_x k_y \right] + E_\ell \sigma_0$ with Fermi velocity $v_\ell$. We find the projected 2D Hamiltonian
\begin{align}\label{eq:hamiltonian_titi}
    \mathcal{H}_{\text{tTI}} =& \sum_{|\vec{k}| < \Lambda} c^\dagger_{\ell,\vec{k}} \begin{pmatrix}h^{\ell = 1}_{R_{-\theta_1}\vec{k}} & T_{0}  \\ T^\dagger_{0}  & h^{\ell = 2}_{R_{-\theta_2}\vec{k}} \end{pmatrix}_{\ell,\ell'} \hspace{-0.5em} c^\pdagger_{\ell',\vec{k}}\\ \nonumber
    &+ \left[ \sum_{|\vec{k}| < \Lambda}\sum_{\pm} c^\dagger_{1,\vec{k}} T_\pm c^\pdagger_{2,\vec{k} \pm \delta\vec{b}} + \text{H.c.} \right]
\end{align}
where $T_0=a_0\sigma_0+ia_z\sigma_z$ and $T_\pm=b_0\pm b_x\sigma_x\pm b_y\sigma_y + i b_z\sigma_z$ are the psudeospin space equivalents to $T_0$ and $T_1$ of \secref{sec:Expl_calc_NINI}. 
The structure of $T_{0,\pm}$ with $a_j, b_j \in \mathbb{R}$ follows when imposing $C_{2z}$ and time-reversal symmetry. Due to the 2D nature of the surface states of the TIs there is no hopping along the vertical direction.
In the numerics we set $v_1=1$ and define the energy scale $\db v_1$ in which we measure all energy scales in this subsection. The energy shifts are chosen as $E_1=0$ and $E_2=-0.7$. 

\begin{figure}[bt]
    \centering
    \includegraphics[width=\linewidth]{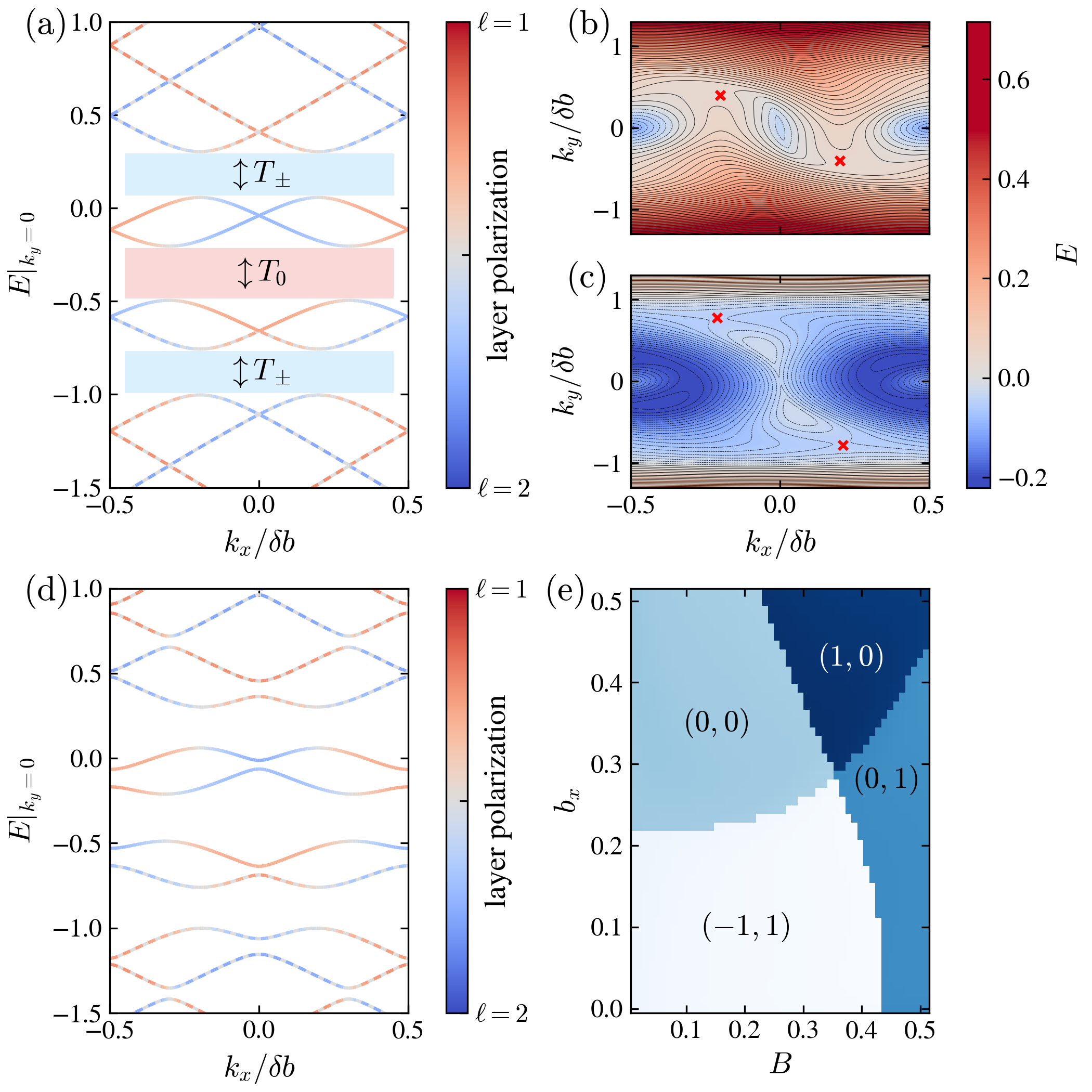}
    \caption{(a) shows the band structure of twisted TIs with Dirac cones at $\Gamma$ in both materials. The color of the bands indicates the layer polarization of the corresponding wave function. The shaded region indicate how the interface interaction terms create isolated bands. The isolated bands at the Fermi energy are represented as a heatmap in panel (b) (upper band) and (c) (lower band). The red crosses indicate the location of Van Hove singularities. Panel (d) shows the Dirac cones gapping out by an additional Zeeman term. The resulting Chern numbers are captured in the phase diagram in panel (e). The labels correspond to the Chern numbers $(C_0, C_1)$.}
    \label{fig:TiTi}
\end{figure}

Figure~\ref{fig:TiTi}(a) shows the resulting 1D band structure at $k_y=0$ as a function of $k_x$. The color indicates the layer polarization. Without any interlayer couplings at the interface, the band structure consists of a red ($\ell=1$) and a blue ($\ell=2$) Dirac cone which are reconstructed in the $\db$-periodic Brillouin zone strip. As our framework describes the generic case of two different materials twisted against each other, a non-vanishing energetic shift of the Dirac cones ($E_1 - E_2 \neq 0$) is to be expected. This leads to an additional intersection of a red and a blue band. However these crossings are avoided due to the $T_0=0.5\sigma_0$ term which open up a gap indicated by the red shaded area in \figref{fig:TiTi}(a). The center of the gap lies at $(E_1 + E_2)/2$, around which the spectrum is particle-hole symmetric.

The $T_\pm=0.4\sigma_0$ term realizes interactions between the momentum quantum numbers $\vec{k}$ and $\vec{k}\pm\vdb$. In \figref{fig:TiTi}(a) this corresponds to avoided crossings between bands which have been folded once ($\vec{k}\pm \vdb$) with unreconstructed bands ($\vec{k}$). This leads to two detached bands on either side of the separated spectrum and is indicated as the blue shaded area in \figref{fig:TiTi}(a). We therefore find a tuneable band-flattening mechanism where the isolated bands are squished from above and below by the $T_0$ and $T_\pm$ terms. The exact nature of the terms \ie $\sigma_{x,y,z}$ contributions do not change this mechanism qualitatively. We further note that the two pairs of bands that are being squished still have Dirac cones at $\vec{k}=(0,0)$ and $(0.5/\delta b,0)$, which are protected by the product of time-reversal and $C_{2z}$.

It is important to keep in mind that the bands are only flattened along the (or close to the) $k_y=0$ cut through the Brillouin zone. To investigate the 2D momentum dependence, we focus on the two isolated bands at the Fermi energy and label them as $B_0$ (lower band) and $B_1$ (upper band). Additionally we define $B_{-1}$ as the band under $B_{0}$, \ie the upper band of the other isolated band pair in \figref{fig:TiTi}(a). To emphasize those three bands, we plot them as solid lines while all other bands are dashed.

Figure~\ref{fig:TiTi}(b) and (c) show contour plots of $B_1$ and $B_0$, respectively. We note that the bands curve up and increase linearly for big values of $k_y$. Since $h_{\vec{k}}^\ell$ scales with $|\vec{k}|$ the contributions of the interface are significant for small $k_y$ and the sign of the band gradient highly depends on $\vec{k}$. This combination of bands curving up in the $k_y$ direction and a highly non-trivial structure around $k_y=0$ favors the appearance of saddle points and thus van Hove singularities in the band structure. We indicate them using  red crosses in \figref{fig:TiTi}(b) and (c). Furthermore $B_0$ shows minima located at $k_y=0$. This is already visible in \figref{fig:TiTi}(a).

Figure~\ref{fig:TiTi}(d) shows how the two Dirac cones at $\vec{k}=(0,0)$ and $\vec{k}=(\pm\frac{\db}{2},0)$ can be gapped out by adding an out of plane Zeeman term to the pseudospin space Hamiltonians in both layers \ie $h_{\vec{k}}^\ell \rightarrow h_{\vec{k}}^\ell + gB\sigma_z$, with $gB=0.1$. This also leads to a non-vanishing Berry curvature in the considered bands. As all bands become linear for $\vert k_y\vert \gg\db$ we do not expect any Berry curvature far from the origin. This is confirmed by the numerics. Thus there is a well-defined Chern number, even though the BZ is not bounded. We calculate it by integrating the Berry curvature over the region $\vec{k}\in [-\db/2, \db/2]\times[-2\db, 2\db]\subset \mathbb{R}^2$. Outside of this region the contributions are effectively zero. In \figref{fig:TiTi}(e), we show a phase diagram constructed by varying the magnetic field $gB$ and the $\sigma_x$ contribution in the $T_\pm$ terms, \ie $b_x$. The resulting phases are characterized by the Chern numbers $(C_0, C_1)$ of the $B_0$ and $B_1$ bands, respectively. At the phase boundaries the gap closes and the system undergoes a topological phase transition. The parameter region was chosen such that band-crossings only occur between $B_{-1}$, $B_0$ and $B_1$. Thus the sum $\sum_{i=-1}^1C_i=1$ is conserved in the considered region, which fixes the Chern number of $B_{-1}$ to $C_{-1}=1-C_0-C_1$. At the boundary of the $(-1,1)$ and the $(0,1)$ phase, $B_{-1}$ and $B_0$ cross and and the respective Chern numbers change according to $C_{-1}=0\rightarrow -1$ and $C_{0}=-1\rightarrow0$. Through this transition $B_1$ stays isolated and thus $C_1$ remains unchanged.

\subsection{Breaking of $C_{2z}$ symmetry at $(\Gamma,K)$}\label{sec:c2z_breaking_gamma_K}
So far we have only considered the first row in \tableref{table:SquareAndTriangularLattice}. Now we want to investigate a twisted heterostructure with electron pockets around the $\Gamma$ ($\ell=1$, square) and $K$ ($\ell=2$, triangular) points. This corresponds to the third row of \tableref{table:SquareAndTriangularLattice}. As emphasized in \secref{SquareTriangularLattice}, we will obtain two quasi-1D channels corresponding to the hybridization across the moir\'e interface of $(\Gamma,K)$ and $(\Gamma, K')$; each such channel does not exhibit $C_{2z}$ symmetry, but the two channels are related by it.

Compared to the $(\Gamma,\Gamma)$ cases above, now there is no $Q=0$ contribution which translates to $T_0=0$. Furthermore, there is only one orientation of $\vdb$ in each of the two channels and not $\pm \vdb$ as before. This leads to the aforementioned broken $C_{2z}$ symmetry and manifests itself in the Hamiltonian of the NIs by not summing over $\pm$ in the second term of \equref{eq:hamiltonian_nini}, i.e., only accounting for the $+$ term in the $(\Gamma,K)$ channel. For the TIs this means $T_-=0$. To look at $C_{2z}$ related channel, $(\Gamma, K')$, one needs to reverse the roles of $T_{+}$ and $T_{-}$, i.e., make the $+$ term vanish and keep the $-$ term.

\begin{figure}[tb]
     \centering
     \includegraphics[width=\linewidth]{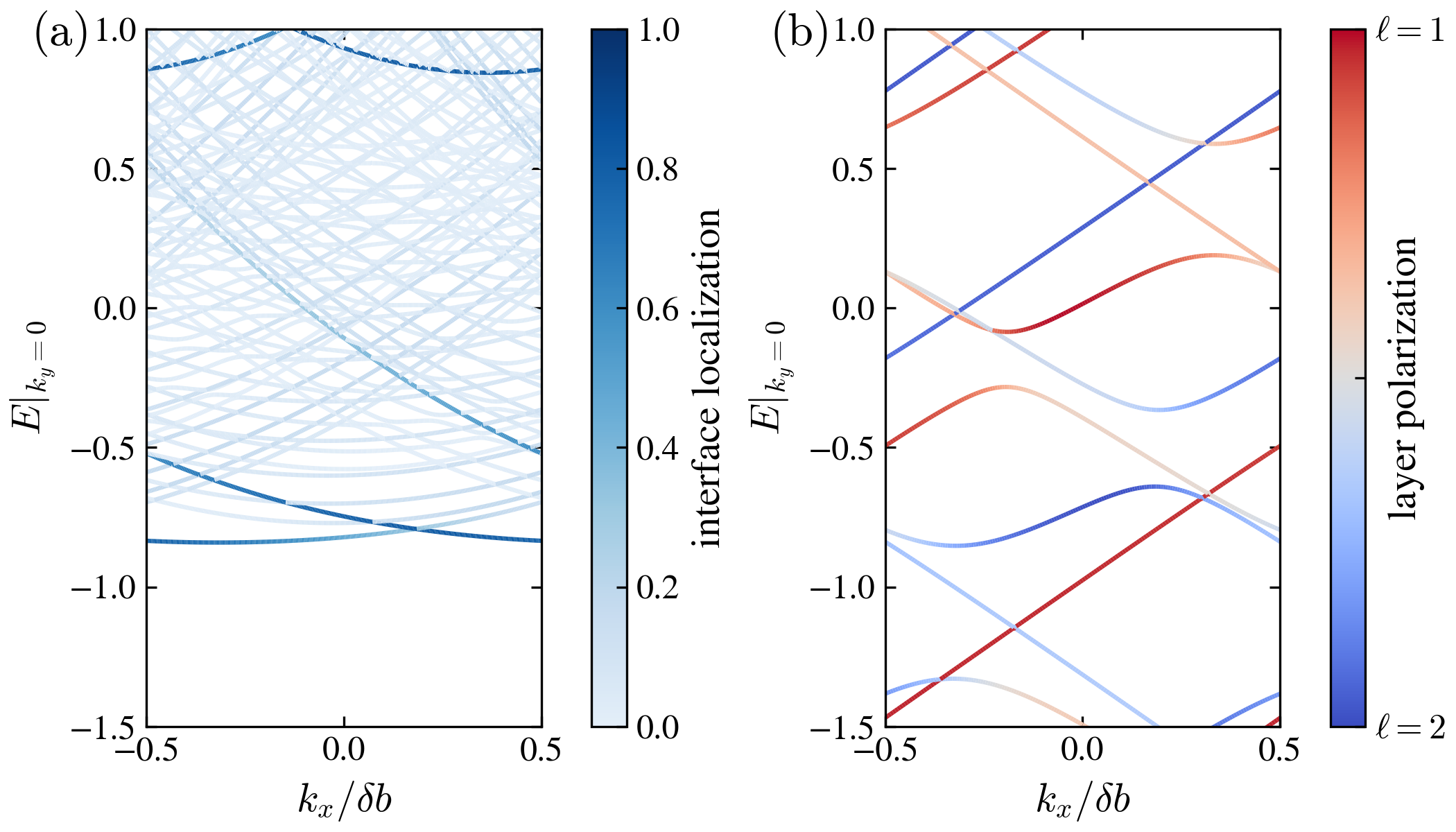}
     \caption{Band structure of twisted 3D NI (a) and twisted 2D TI surface states (b) with electron pockets around $(\Gamma, K)$ as a function of $k_x$ along the $k_y=0$ cut. The coloring of the bands indicates the interface localization (a) and layer porlaization (b) of the corresponding wave functions.}
     \label{fig:K-Gamma}
 \end{figure}

Figure~\ref{fig:K-Gamma} shows the band structures in the $(\Gamma,K)$ channel for twisted NIs (a) and TIs (b) at $k_y=0$. All energies are given in units of $v_1\db$.
When comparing \figref{fig:NiNi}(b) and \figref{fig:K-Gamma}(a), we note that the lowest energy band is still localized at the surface, however it is not isolated in the latter plot. This is due to the absence of the $T_0$ term which leads to a band repulsion between the lowest energy bands of both materials. Due to the broken $C_{2z}$ symmetry the $k_y=0$ band structure is not symmetric as a function of $k_x$. 

A comparison of \figref{fig:TiTi}(a) and \figref{fig:K-Gamma}(b) also clearly shows the broken $C_{2z}$ symmetry in the $(\Gamma, K)$ plot. Furthermore we note that the band-flattening mechanism described in \secref{sec:Expl_calc_TiTi} does not apply for this case since the $T_0$ and $T_-$ terms are zero. As a consequence, the band structure is only separated at the crossing of the unfolded $\ell=1$ and one-time-folded $\ell=2$ bands.
Figure~\ref{fig:K-Gamma}(a) was obtained by setting $T_1=0.7$, $\mu_1=0$, $\mu_2=0.2$ and $m_2^*/m_1^*=1.3$. For \figref{fig:K-Gamma}(b) we used $T_+=0.4\sigma_0+0.2\sigma_x+0.2\sigma_y+i0.1\sigma_z$ and $E_1=0$ $E_2=-0.7$. However the broken symmetry is a feature of the model and can be reproduced with rather arbitrary parameters.

In the examples above, we investigated interface superlattices with equal properties, \ie NI-NI and TI-TI. Upon studying different combinations, we found that similar band structures in both materials typically lead to more transparent band features as there is less complexity. Nevertheless to further demonstrate the diversity of opportunities at moiré heterostructures and that our developed framework can be applied to arbitrary combinations of types of band structures, we discuss the 1D physics arising from a twisted TI with 2D surface states at the interface with a 3D NI in \appref{sec:app_additioan_band_struc}.

\subsection{Geometric magic angles}
We finally note that the vanishing of the momentum transfer $\delta \vec{b}$ in \equref{DefinitionOfdeltab} at non-trivial crystalline reference points with finite twist angles leads to a novel type of ``magic angle''. As opposed to the magic angle of, e.g., twisted bilayer graphene \cite{bistritzer2011moire}, it is not an interference effect of intralayer propagation through the moiré unit cell and interlayer tunneling processes that leads to flat bands. Instead, it is the result of the different momentum-space locations $\vec{P}_{j,\ell}$ of the low-energy electronic degrees of freedom in the two \textit{distinct} materials on either side of the moiré interface that lead to the vanishing of $\delta \vec{b}$ at non-trivial twist angles. As the ratio of the tunneling energy scales and intralayer propagation does not determine the twist angle $\theta^*$ of the reference point but instead the purely geometric condition in \equref{DefinitionOfdeltab}, we refer to the non-zero angles of the leading crystalline reference points as \textit{geometric magic angles}. We note that, of course, there are also ``magic lattices mismatches'', but $\eta$ is typically harder to control experimentally than the twist angle.

\begin{figure}[tb]
     \centering
     \includegraphics[width=\linewidth]{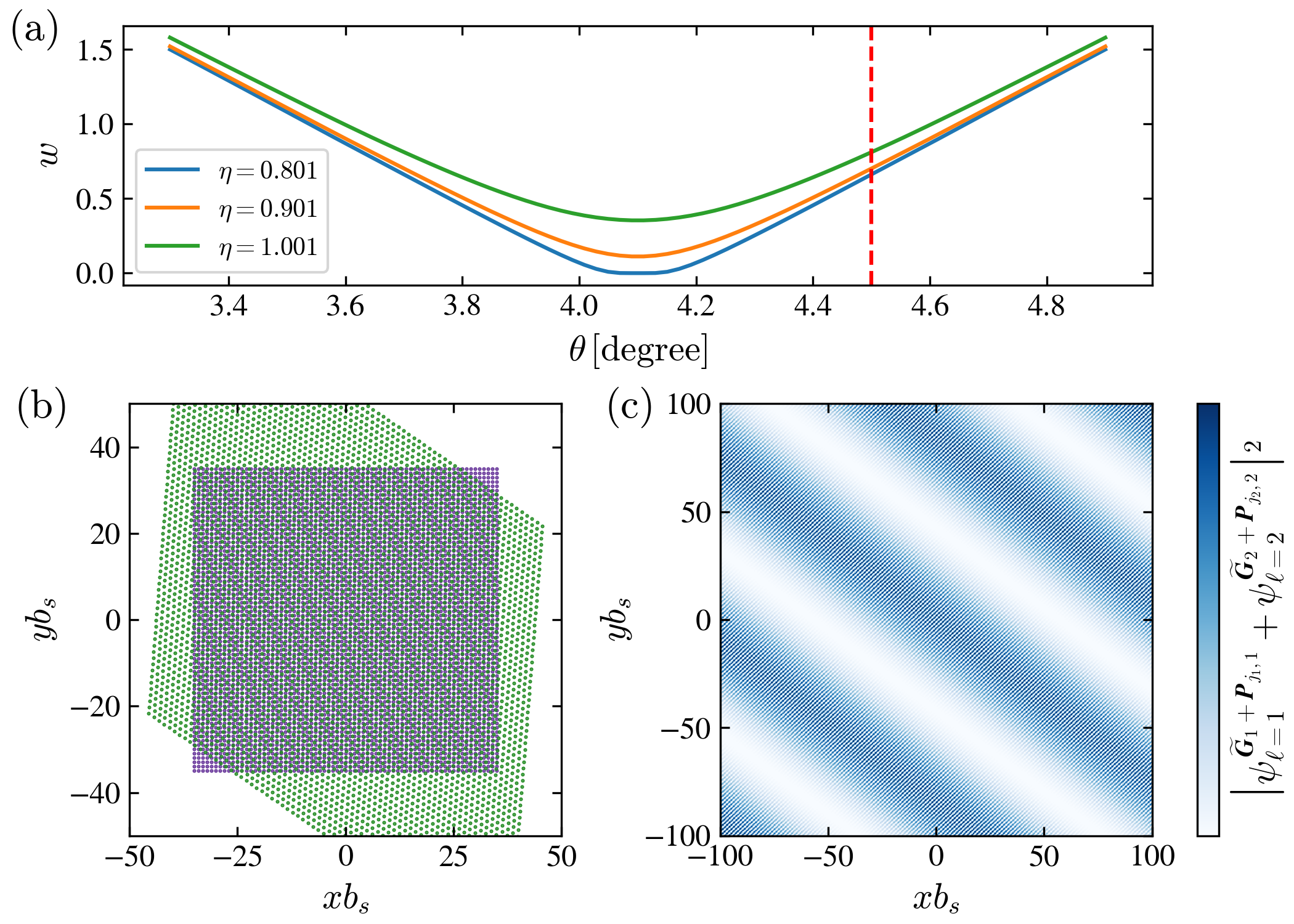}
     \caption{Bandwidth $w$ as a function of twist angle $\theta$ for different lattice mismatches $\eta$ calculated for twisted TIs with Dirac cones at $M$ (square) and $\Gamma$ (triangle) (a). The plot is centered around the crystalline reference point corresponding to the third row in \tableref{table:SquareAndTriangularLattice}, \ie $(\theta^*, \eta^*)\approx(4.1^{\circ}, 0.801).$ The red, dashed line indicates the twist angle $\theta=4.5^{\circ}$ for which the real space lattices (triangular in green, square in purple) are plotted in panel (b) at $\eta = \eta^*$. Panel (c) shows the interference of the corresponding Bloch waves at the same lattice mismatch and twist.}
     \label{fig:geometricmagicangles}
 \end{figure}

To illustrate this further, we consider in \figref{fig:geometricmagicangles} the example of square and triangular lattices with low-energy bands around $M$ and $\Gamma$, respectively. We can read off from \tableref{table:SquareAndTriangularLattice} that the leading crystalline reference point is at $\theta^* \approx 4.1^\circ$ and $\eta^* \approx 0.801$. 
As discussed in more detail in \secref{sec:c2z_breaking_gamma_K}, the corresponding interlayer Hamiltonian does not include a $T_0$ term. The resultant band structure is similar to \figref{fig:TiTi}(a) without the gap indicated by the red box, and features four gapless low energy bands which are isolated from the rest of the spectrum. We therefore define $w$ as the combined bandwidth of those four bands.
As can be seen in \figref{fig:geometricmagicangles}(a) the bands become very flat (along the direction of $\delta \vec{b}$) for $\theta$ close to $\theta^*$. It is important to note that the \textit{Bravais lattices}' moiré pattern does not become very large close to this point, see \figref{fig:geometricmagicangles}(b); the flattening of the bands is rather a consequence of the spatial interference of the \textit{Bloch waves}. To illustrate this, we show in \figref{fig:geometricmagicangles}(c) the spatial interference of Bloch waves, defined as $\psi_\ell^{\widetilde{\vec{G}}_\ell  + \vec{P}_{j,\ell}}(\vec{r})\propto e^{i (\widetilde{\vec{G}}_\ell + \vec{P}_{j,\ell})\cdot \vec{r}}$, associated with $\Gamma$ and $M$ points of the square and triangular lattice at the same twist angle as in (b) which is close to $\theta^*$. Here, $(\widetilde{\vec{G}}_1, \widetilde{\vec{G}}_2)$ is the pair of reciprocal lattice vectors yielding minimal $\vdb$ according to \equref{DefinitionOfdeltab}. In this case, indeed, we see a long-wave length moiré modulation, which diverges to infinity right at the continuum reference point~\footnote{See also animation available as supplementary movie file.}.

\section{Conclusion and Outlook}
In this work, we studied the electronic spectrum at an interface between two different materials, with in general different crystal structures and Bravais lattices, as shown in \figref{fig:introfig}. For a given set of high-symmetry momentum points $\vec{P}_{j,\ell}$ around which the low-energy degrees of freedom in the Brillouin are localized in each material $\ell =1,2$, we studied the sets of lattice parameters $p^*$, i.e., the relative orientation and length of the primitive vectors of the lattices, for which the electronic spectrum can be approximated as being crystalline. Such a ``crystalline reference point'' $p^*$ is determined by the existence of reciprocal lattice vectors $\vec{G}_j$ such that $\delta b \rightarrow 0$ in \equref{DefinitionOfdeltab}, when the lattice parameters approach $p^*$. Each crystalline reference point is characterized by a momentum $Q$, defined as the smallest possible value of \equref{DefinitionOfQGen}; the larger Q, the weaker the associated moir\'e-reconstruction of the bands. The, in this sense, leading set of continuum reference points are summarized in \tableref{table:SquareAndTriangularLattice} for each combination of high symmetry points $\vec{P}_{j,\ell}$ for one material exhibiting a square and the other a triangular lattice in the respective lattice planes forming the interface. We see that, as consequence of the two lattices being different, some of the continuum reference points are not at zero twist (and lattice mismatch). We further observe that a variety of different types of band reconstruction is possible, which includes (i) a single quasi-1D channel with time-reversal and $C_{2z}$ symmetry ($\Gamma$-$\Gamma$ and $M$-$\Gamma$), (ii) two 1D channels, related by time-reversal/$C_{2z}$ symmetry ($\Gamma$-$K$ and $M$-$K$), and (iii) mixed dimensionality (all other combinations) in the sense that a subset of the pockets exhibit a significant moir\'e modulation leading to quasi-1D physics (with one or two channels) while the complement is effectively unreconstructed and, thus, 2D.

While the generalization to other pairs of Bravais lattices is straightforward, for our explicit band structure calculations, we focused on square and triangular lattices meeting at the interface. We considered (a) quadratic bulk bands in both systems, (b) an interface between a topological insulator and bulk materials with a quadratic band, and (c) two topological insulators, uncovering a variety of interesting features. For instance, in case (a), which is also directly relevant to the experiment in \refcite{MannhartExperiment}, we observe that the presence or absence of a well-defined interface band, i.e., a band that is both energetically separated (at least in part of the Brillouin zone) from the other bulk bands and localized in the vicinity of the interface, strongly depends on the nature of bulk bands: if the bands on either side are both particle-like or both hole-like, there is a regime where the moir\'e reconstruction leads to a well-defined interface band, even when the materials have sizable hopping processes $t_z$ between the lattice planes; if one band is particle- and the other hole-like, the surface bands will quickly hybridize and be absorbed by the bulk states when $t_z$ is turned on. To mention a second example, we show that twisting two different TIs, one with square and one with a triangular Bravais lattice in the relevant lattice planes, gives rise to a quasi-1D surface theory with Dirac cones (even when both materials, considered separately, have surface Dirac cones at the $\Gamma$ point) located at the time-reversal invariant points $(0,0)$ and $(\pi,0)$; these are gapped out by a Zeeman field that can induce multiple topological phase transitions. 

Naturally, there are many open questions concerning the physics of moir\'e interfaces. First, using the formalism of this work, there is a plethora of different combinations of lattices, high-symmetry points, and types of bands in the two materials to explore. Second, an important extension of the current work would be to include lattice relaxation effects in the vicinity of the interface, which result from the mutual influence of the two incompatible lattices, and study its impact on the electronic properties. Finally, there is a large set of questions associated with the impact of electronic correlations in these heterostructures. For instance, the tunable quasi-1D nature of some of the bands \cite{OurOxideInteraction,PhysRevB.108.L201120} as well as the presence and tunability of van Hove singularities \cite{ChubukovSC,PhysRevX.8.041041} very naturally comes with enhanced correlation effects, likely leading to competing superconducting and particle-hole instabilities, such as charge- or spin-density waves. In that regard, the systems with mixed dimensionality are particularly interesting since the mutual impact of quasi-1D and 2D channels has the potential to induce rich physics.

\begin{acknowledgments}
The authors thank Laura Classen, Varun Harbola, Seungwon Jung, and Jochen Mannhart for fruitful discussions. All authors further acknowledge funding by the European Union (ERC-2021-STG, Project 101040651---SuperCorr). Views and opinions expressed are however those of the authors only and do not necessarily reflect those of the European Union or the European Research Council Executive Agency. Neither the European Union nor the granting authority can be held responsible for them. 
\end{acknowledgments}

%\bibliography{draft_Refs}
\input{draft_2.bbl}

\onecolumngrid

\begin{appendix}

\section{Detailed derivation of momentum-space model}\label{DetailedModelDerivation}
Since the momentum-space model, described briefly in \secref{MomentumSpaceModel}, is a central to our analysis, we here present a detailed derivation. 

\subsection{Two layers of one-band models}
Although we are ultimately interested in twisting 3D bulk crystals with in general multiple relevant bands, we start for pedagogical reasons by considering two quasi-2D materials with only one active band each but allowing for arbitrary and possibly different Bravais lattices in each layer. Once the basic notation is established in this minimal setup, the generalization will be straightforward. 

More specifically, the Hamiltonian we start from has two components $\mathcal{H} = \mathcal{H}_1 + \mathcal{H}_2$, where
\begin{equation}
    \mathcal{H}_1 = \sum_{\ell =1,2} \int \diff^2 \vec{r} \, c^\dagger_{\ell}(\vec{r}) \left[ -\frac{1}{2m_{\ell}} \Delta + \mu_\ell + V_\ell( \mathcal{R}_{-\theta_\ell} \vec{r}) \right] c^\pdagger_{\ell}(\vec{r}). \label{StartingPointHamiltonianInEachLayer}
\end{equation}
Here, $c^\dagger_{\ell}(\vec{r})$ is the electronic creation operator at position $\vec{r} \in \mathbb{R}^2$ and in layer $\ell$, $\mathcal{R}_\theta$ is a 2D rotation matrix by angle $\theta$, $\theta_\ell$ the twist angle, and $V_{\ell}(\vec{r})$ is the (untwisted) lattice potential in layer $\ell$. The latter obeys
\begin{equation}
    V_{\ell}(\vec{r}) = V_{\ell}(\vec{r}+\vec{R}), \qquad \vec{R} \in \text{BL}_\ell,
\end{equation}
defining the Bravais lattice in the untwisted layer $\ell$. We here allow for general $\text{BL}_\ell$ which may or may not differ in the two layers. In \equref{StartingPointHamiltonianInEachLayer}, we neglected spin-orbit coupling, which is why we do not need to include an explicit spin index. 
The second term in the Hamiltonian,
\begin{equation}
    \mathcal{H}_2 = \frac{1}{A}\int \diff^2 \vec{r} \int \diff^2 \vec{r}' \, t(|\vec{r}-\vec{r}'|)c^\dagger_1(\vec{r}) c^\pdagger_2(\vec{r}') + \text{H.c.}, \label{TunnelingHamiltonian}
\end{equation}
describes the interlayer tunneling, where $A$ is the area of the layers.

We start from the Bloch states $\psi_{\vec{k},\ell}(\vec{r})$ of the band of interest of the uncoupled and untwisted layers, obeying 
\begin{equation}
    \left[ -\frac{1}{2m_{\ell}} \Delta + \mu_\ell + V_\ell(\vec{r}) \right]\psi_{\vec{k},\ell}(\vec{r}) = \epsilon_{\vec{k},\ell} \psi_{\vec{k},\ell}(\vec{r}), \quad  \vec{k}\in\text{BZ}_\ell,
\end{equation}
where $\text{BZ}_\ell$ is the first Brillouin zone of the Bravais lattice $\text{BL}_\ell$ (both untwisted). With these states at hand, it is straightforward to project $\mathcal{H}_1$ onto this band and diagonalize it; we set (with $N_\ell$ denoting the number of unit cells in layer $\ell$)
\begin{equation}
    c_\ell(\vec{r}) = \frac{1}{\sqrt{N_\ell}} \sum_{\vec{k}\in\text{BZ}_\ell} \psi_{\vec{k},\ell}(R_{-\theta_\ell}\vec{r}) \bar{c}_{\ell,\vec{k}} \label{cbarDefinition}
\end{equation}
such that, under band projection,
\begin{equation}
    \mathcal{H}_1 \, \rightarrow \, \mathcal{H}^P_1 = \sum_{\ell=1,2} \sum_{\vec{k}\in\text{BZ}_\ell} \epsilon_{\vec{k},\ell} \bar{c}^\dagger_{\ell,\vec{k}} \bar{c}^\pdagger_{\ell,\vec{k}}. \label{IntraTermBeforeTrafo}
\end{equation}
While this result looks very compact, the current choice of basis makes the physics less transparent: as is apparent from \equref{cbarDefinition}, the two different fermion species are defined in different bases that are rotated relative to each other. This will make the tunneling terms, to be considered below, look less clear since momentum conservation will become obscured (e.g., $\bar{c}^\dagger_{1,\vec{k}} \bar{c}^\pdagger_{2,\vec{k}}$ does \textit{not} conserve momentum). To this end, we choose to work in the ``lab frame''. To motivate that basis and for future reference, we introduce Wannier functions $w_\ell(\vec{r}-\vec{R})$ in the usual way,
\begin{equation}
    \psi_{\ell,\vec{k}}(\vec{r}) = \frac{1}{\sqrt{N_\ell}} \sum_{\vec{R} \in \text{BL}_\ell} w_\ell(\vec{r}-\vec{R}) e^{i\vec{k}\vec{R}} \, \,\Leftrightarrow\,\, w_\ell(\vec{r}-\vec{R})  = \frac{1}{\sqrt{N_\ell}} \sum_{\vec{k} \in \text{BZ}_\ell} \psi_{\ell,\vec{k}}(\vec{r}) e^{-i\vec{k}\vec{R}}.
\end{equation}
The associated Wannier functions after rotating the two layers are now just given by
\begin{equation}
    w_{\ell}^{\theta_\ell}(\vec{r}-\vec{R}) := w_\ell( \mathcal{R}_{-\theta_\ell} ( \vec{r}-\vec{R})),
\end{equation}
which we can now use to define Bloch states for both layers in the same ``lab frame'' of reference,
\begin{equation}
    \psi^{\theta_\ell}_{\ell,\vec{k}}(\vec{r}) = \frac{1}{\sqrt{N_\ell}} \sum_{\vec{R} \in \text{BL}_{\ell,\theta_\ell}} w^{\theta_\ell}_\ell(\vec{r}-\vec{R}) e^{i\vec{k}\vec{R}},
\end{equation}
where $\text{BL}_{\ell,\theta_\ell} := \{\mathcal{R}_{\theta_\ell}\vec{R}\,| \vec{R} \in \text{BL}_{\ell} \}$ is the Bravais lattice $\text{BL}_{\ell}$ rotated by $\theta_\ell$. To relate $\psi_{\ell,\vec{k}}(\vec{r})$ and $\psi^{\theta_\ell}_{\ell,\vec{k}}(\vec{r})$, consider
\begin{align}
    \psi_{\ell,\vec{k}}(R_{-\theta_\ell}\vec{r}) &= \frac{1}{\sqrt{N_\ell}} \sum_{\vec{R} \in \text{BL}_\ell} w_\ell(R_{-\theta_\ell}\vec{r}-\vec{R}) e^{i\vec{k}\vec{R}} \\
    &= \frac{1}{\sqrt{N_\ell}} \sum_{\vec{R}' \in \text{BL}_{\ell,\theta_\ell}} w_\ell(R_{-\theta_\ell}(\vec{r}-\vec{R}')) e^{i\vec{k}(R_{-\theta_\ell} \vec{R}')} \\
    &= \frac{1}{\sqrt{N_\ell}} \sum_{\vec{R}' \in \text{BL}_{\ell,\theta_\ell}} w^{\theta_\ell}_\ell(\vec{r}-\vec{R}') e^{i(R_{\theta_\ell}\vec{k})\vec{R}'} = \psi^{\theta_\ell}_{\ell,R_{\theta_\ell}\vec{k}}(\vec{r}).
\end{align}
This allows us to rewrite \equref{cbarDefinition} as
\begin{equation}
    c_\ell(\vec{r}) = \frac{1}{\sqrt{N_\ell}} \sum_{\vec{k}\in\text{BZ}_\ell} \psi_{\vec{k},\ell}(R_{-\theta_\ell}\vec{r}) \bar{c}_{\ell,\vec{k}} = \frac{1}{\sqrt{N_\ell}} \sum_{\vec{k}\in\text{BZ}_\ell} \psi^{\theta_\ell}_{R_{\theta_\ell}\vec{k},\ell}(\vec{r}) \bar{c}_{\ell,\vec{k}} \equiv \frac{1}{\sqrt{N_\ell}} \sum_{\vec{k}'\in\text{BZ}_{\ell,\theta_\ell}} \psi^{\theta_\ell}_{\vec{k}',\ell}(\vec{r}) c_{\ell,\vec{k}'},
\end{equation}
where we introduced the new operators 
\begin{equation}
    c_{\ell,\vec{k}} = \bar{c}_{\ell,R_{-\theta_\ell}\vec{k}} \label{DifferentFrame}
\end{equation}
and the rotated Brillouin zone $\text{BZ}_{\ell,\theta_\ell}:=\{\mathcal{R}_{\theta_\ell}\vec{k}\,| \vec{k} \in \text{BZ}_{\ell} \}$, i.e., the Brillouin zone associated with the rotated Bravais lattice $\text{BL}_{\ell,\theta_\ell}$.

Using these operators, the projected intralayer Hamiltonian becomes
\begin{equation}
    \mathcal{H}^P_1 = \sum_{\ell=1,2} \sum_{\vec{k}\in\text{BZ}_{\ell,\theta_\ell}} \epsilon_{R_{-\theta_\ell}\vec{k},\ell} c^\dagger_{\ell,\vec{k}} c^\pdagger_{\ell,\vec{k}}. \label{HP1SimplestCase}
\end{equation}
This form makes sense in that, in the lab frame, the dispersion should rotate with the lattice. We will next see that also the tunneling part of the Hamiltonian is more intuitive in this bases. 

To project $\mathcal{H}_2$ in \equref{TunnelingHamiltonian} onto the band of interest, we write
\begin{equation}
    c_\ell(\vec{r}) = \frac{1}{\sqrt{N_\ell}} \sum_{\vec{k}\in\text{BZ}_{\ell,\theta_\ell}} \psi^{\theta_\ell}_{\vec{k},\ell}(\vec{r}) c_{\ell,\vec{k}} = \frac{1}{N_\ell} \sum_{\vec{k}\in\text{BZ}_{\ell,\theta_\ell}} c_{\ell,\vec{k}} \sum_{\vec{R} \in \text{BL}_{\ell,\theta_\ell}} w^{\theta_\ell}_\ell(\vec{r}-\vec{R}) e^{i\vec{k}\vec{R}}
\end{equation}
which we insert into $\mathcal{H}_2$, yielding
\begin{align}\begin{split}
    \mathcal{H}_2 \, \rightarrow \, \mathcal{H}^P_2 &= \frac{1}{N_1N_2} \sum_{\vec{k}_{1}\in\text{BZ}_{1,\theta_1} } \sum_{\vec{k}_{2}\in\text{BZ}_{2,\theta_2} } c^\dagger_{1,\vec{k}_1} c^\pdagger_{2,\vec{k}_2} \sum_{\vec{R}_1 \in \text{BL}_{1,\theta_1}} \sum_{\vec{R}_2 \in \text{BL}_{2,\theta_2}} e^{i (\vec{k}_2\vec{R}_2 - \vec{k}_1\vec{R}_1)} \\
    & \qquad \quad  \times \frac{1}{A} \int \diff^2 \vec{r} \int \diff^2 \vec{r}' \, (w^{\theta_1}_1(\vec{r}-\vec{R}_1))^* t(|\vec{r}-\vec{r}'|) w^{\theta_2}_2(\vec{r}'-\vec{R}_2)  + \text{H.c.}. \label{RewritingH2}
\end{split}\end{align}
We now define
\begin{equation}
    \tilde{t}(\vec{R}_1-\vec{R}_2) := \frac{1}{A}\int \diff^2 \vec{r} \int \diff^2 \vec{r}' \, (w^{\theta_1}_1(\vec{r}-\vec{R}_1))^* t(|\vec{r}-\vec{r}'|) w^{\theta_2}_2(\vec{r}'-\vec{R}_2) =: \sum_{\vec{q}} T_{\vec{q}} e^{i\vec{q}(\vec{R}_1-\vec{R}_2)}, \label{FourierTransformOfCoupling}
\end{equation}
leading to
\begin{align}\begin{split}
    \mathcal{H}^P_2 &= \frac{1}{N_1N_2} \sum_{\vec{k}_{1}\in\text{BZ}_{1,\theta_1} } \sum_{\vec{k}_{2}\in\text{BZ}_{2,\theta_2} } c^\dagger_{1,\vec{k}_1} c^\pdagger_{2,\vec{k}_2} \sum_{\vec{R}_1 \in \text{BL}_{1,\theta_1}} \sum_{\vec{R}_2 \in \text{BL}_{2,\theta_2}} \sum_{\vec{q}} T_{\vec{q}} e^{i ((\vec{k}_2-\vec{q})\vec{R}_2 - (\vec{k}_1-\vec{q})\vec{R}_1)} + \text{H.c.}.
\end{split}\end{align}
We next use that
\begin{equation}
    \frac{1}{N_\ell} \sum_{\vec{R} \in \text{BL}_{\ell,\theta_\ell}} e^{i\vec{k}\vec{R}} = \sum_{\vec{G} \in \text{RL}_{\ell,\theta_\ell}} \delta_{\vec{k},\vec{G}} =: \delta_{\vec{k} \in \text{RL}_{\ell,\theta_\ell}}, \label{RLIdentity}
\end{equation}
where $\text{RL}_{\ell,\theta_\ell}$ is the reciprocal lattice of $\text{BL}_{\ell,\theta_\ell}$.
This leads to the two closely related final forms of the interlayer part of the Hamiltonian
\begin{subequations}\begin{align}
    \mathcal{H}^P_2 &= \sum_{\vec{k}_{\ell}\in\text{BZ}_{\ell,\theta_\ell} }  c^\dagger_{1,\vec{k}_1} c^\pdagger_{2,\vec{k}_2} \sum_{\vec{q}} T_{\vec{q}} \delta_{\vec{k}_1-\vec{q} \in \text{RL}_{1,\theta_1}} \delta_{\vec{k}_2-\vec{q} \in \text{RL}_{2,\theta_2}} + \text{H.c.} \\
    &= \sum_{\vec{k}_{\ell}\in\text{BZ}_{\ell,\theta_\ell} }  \sum_{\vec{G}_\ell \in \text{RL}_{\ell,\theta_\ell}} c^\dagger_{1,\vec{k}_1} c^\pdagger_{2,\vec{k}_2}  \delta_{\vec{k}_1 +\vec{G}_1,\vec{k}_2 +\vec{G}_2} T_{\vec{k}_1 +\vec{G}_1}  + \text{H.c.}. 
\end{align}
\label{FormOfH2P}\end{subequations}
The total momentum-space Hamiltonian, projected to the band of interested, is then given by $\mathcal{H}^P_1 + \mathcal{H}^P_2$ with $\mathcal{H}^P_1$ and $\mathcal{H}^P_2$ given by \equsref{HP1SimplestCase}{FormOfH2P}, respectively. 

\subsection{Generalization to multiple bands and bulk materials}
As anticipated above, we now generalize this form to multiple active bands and taking into account that the materials between which an interface is formed are not quasi 2D but instead 3D bulk materials. We start by considering each material $\ell =1,2$ separately and denote their respective (for now untwisted) Bloch Hamiltonian as $h^\ell(-i \hbar \vec{\nabla},\vec{r},z)$, where $z \in \mathbb{R}$ is the direction perpendicular to the interface that we will consider below and $\vec{r}\in \mathbb{R}^2$ are the remaining two orthogonal directions. The 3D Bravais-lattice translational symmetry of each material is expressed as
\begin{equation}
    h_{\text{B}}^\ell(-i \hbar \vec{\nabla},\vec{r},z) = h_{\text{B}}^\ell(-i \hbar \vec{\nabla},\vec{r} + \vec{R},z) = h_{\text{B}}^\ell(-i \hbar \vec{\nabla},\vec{r},z + Z), \qquad \vec{R} \in \text{BL}_\ell, \quad Z/d_\ell \in \mathbb{Z},
\end{equation}
i.e., $Z$ parametrizes the members of a family of lattice planes with distance $d_\ell$; within each such plane, the system has discrete translational symmetry leading to the 2D Bravais lattice $\text{BL}_\ell$. Note that $h_{\text{B}}^\ell$ is a matrix in spin space and can include spin-orbit coupling. 

Let us assume that only a subset of $N_b^\ell$ bands of $h_{\text{B}}^\ell(-i \hbar \vec{\nabla},\vec{r},z)$ are relevant for the phenomena of interest and let us include enough bands such that there are no obstructions to constructing symmetric, exponentially localized Wannier functions $w_{\ell,\alpha}(\vec{r}-\vec{R},z-Z)$, $\alpha = 1,2,\dots N_b^\ell$; we denote their centers by $(\vec{R},Z) + (\vec{d}_\alpha,z_\alpha)$. In the absence of spin-orbit coupling (the presence of inversion symmetry), $\alpha$ can be viewed as a multi-index, labeling combinations of spin-$\uparrow$/$\downarrow$ (pseudospin-$\uparrow$/$\downarrow$) and the different bands. Twisting the crystal along the $z$ direction by $\theta_\ell$, the Wannier functions transform as 
\begin{equation}
    w_{\ell,\alpha}(\vec{r}-\vec{R},z-Z) \, \longrightarrow \,  w_{\ell,\alpha}(\mathcal{R}_{-\theta_\ell}\vec{r}-\vec{R},z-Z).
\end{equation}
which are now localized around the rotated centers $(\vec{R}',Z) + (\mathcal{R}_{\theta_\ell}\vec{d}_\alpha,z_\alpha)$ where $\vec{R}'\in \text{BL}_{\ell,\theta_\ell}$ [cf.~also \equref{cbarDefinition} for the same transformation of the Bloch states]. We denote the electronic creation operators for these basis states by $a^\dagger_{\ell,\vec{R},Z,\alpha}$ which allows us to state the Hamiltonian of the heterostructure involving both materials and the interface: as before, the Hamiltonian consists of two parts, $\mathcal{H}_{\text{bulk}} + \mathcal{H}_{\text{inter}}$, with 
\begin{equation}
    \mathcal{H}_{\text{bulk}} = \sum_{\ell =1 ,2} \sum_{\vec{R},\vec{R}' \in \text{BL}_\ell} \sum_{Z,Z' \in \mathcal{Z}_\ell} \sum_{\alpha,\alpha'=1}^{N_b^\ell} a^\dagger_{\ell,\vec{R},Z,\alpha} \hat{h}_{\alpha,\alpha'}^\ell(\vec{R}-\vec{R}',Z,Z') a^\pdagger_{\ell,\vec{R}',Z',\alpha'}, \label{BulkHamiltonian}
\end{equation}
where $\hat{h}^\ell$ contains all the tight-binding matrix elements and $\mathcal{Z}_\ell$ is the set of lattice planes of material $\ell$; for concreteness, we will choose the interface between the topmost layer ($Z=0$) of $\ell = 1$ and the bottom layer ($Z=d_2$) of $\ell = 2$. We then have $\mathcal{Z}_1 = \{0,-d_1,-2d_1,\dots \}$ and $\mathcal{Z}_2 = \{d_2, 2d_2, 3d_2,  \dots \}$. The coupling between the two materials across the interface is captured by
\begin{equation}
    \mathcal{H}_{\text{inter}} = \frac{1}{\sqrt{N_1N_2}}\sum_{\vec{R}_\ell \in \text{BL}_\ell} \sum_{\alpha_\ell=1}^{N_b^\ell} a^\dagger_{1,\vec{R}_1,Z=0,\alpha_1} \bar{t}_{\alpha_1,\alpha_2}\hspace{-0.3em}\left(\mathcal{R}_{\theta_1}(\vec{R}_1 + \vec{d}_{\alpha_1}) - \mathcal{R}_{\theta_2}(\vec{R}_2 + \vec{d}_{\alpha_2}),z_{\alpha_1}-(d+z_{\alpha_2})\right)a^\pdagger_{2,\vec{R}_2,Z=d_2,\alpha_2} + \text{H.c.},
\end{equation}
where we made the natural assumption %[in line with \equref{FourierTransformOfCoupling} above] 
that only the degrees of freedom on the layers closest to the interface couple significantly and that the coupling strength is a function of the distance between the respective Wannier centers and of $\alpha_{1,2}$ (the type of Wannier state). To keep the notation compact, we write
\begin{equation}
    t_{\alpha_1,\alpha_2}(\mathcal{R}_{\theta_1}\vec{R}_1 - \mathcal{R}_{\theta_2}\vec{R}_2) \equiv \bar{t}_{\alpha_1,\alpha_2}\hspace{-0.1em}(\mathcal{R}_{\theta_1}(\vec{R}_1 + \vec{d}_{\alpha_1}) - \mathcal{R}_{\theta_2}(\vec{R}_2 + \vec{d}_{\alpha_2}),z_{\alpha_1}-(d+z_{\alpha_2}))
\end{equation}
in the following. 

We continue by rewriting $\mathcal{H}_{\text{bulk}}$ via introduction of partially Fourier-transformed operators,
\begin{equation}
    a_{\ell,\vec{R},Z,\alpha} = \frac{1}{\sqrt{N_\ell}} \sum_{\vec{k}\in \text{BZ}_\ell} e^{i \vec{k} \vec{R}} \bar{c}_{\ell,\vec{k},Z,\alpha}, \label{FourierTransformOperators}
\end{equation}
such that \equref{BulkHamiltonian} becomes
\begin{equation}
    \mathcal{H}_{\text{bulk}} = \sum_{\ell =1 ,2} \sum_{\vec{k}\in \text{BZ}_\ell} \sum_{Z,Z' \in \mathcal{Z}_\ell} \sum_{\alpha,\alpha'=1}^{N_b^\ell} \bar{c}^\dagger_{\ell,\vec{k},Z,\alpha} h_{\alpha,\alpha'}^\ell(\vec{k},Z,Z') \bar{c}^\pdagger_{\ell,\vec{k},Z',\alpha'}, \label{BulkBandHam}
\end{equation}
where $h_{\alpha,\alpha'}^\ell(\vec{k},Z,Z') = \sum_{\vec{R}\in\text{BL}_\ell} e^{i\vec{k}\vec{R}} \hat{h}_{\alpha,\alpha'}^\ell(\vec{R},Z,Z')$. Equation (\ref{BulkBandHam}) is the analogue of \equref{IntraTermBeforeTrafo}; similar to our analysis above [cf.~\equref{DifferentFrame}], we will transform to the `lab frame',
\begin{equation}
     c_{\ell,\vec{k},Z,\alpha} = \bar{c}_{\ell,\mathcal{R}_{-\theta_\ell}\vec{k},Z,\alpha}, \label{SecondTransformationLabFrame} 
\end{equation}
and arrive at
\begin{equation}
    \mathcal{H}_{\text{bulk}} = \sum_{\ell =1 ,2} \sum_{\vec{k}\in \text{BZ}_{\ell,\theta_\ell}} \sum_{Z,Z' \in \mathcal{Z}_\ell} \sum_{\alpha,\alpha'=1}^{N_b^\ell} c^\dagger_{\ell,\vec{k},Z,\alpha} h_{\alpha,\alpha'}^\ell(\mathcal{R}_{-\theta_\ell}\vec{k},Z,Z') c^\pdagger_{\ell,\vec{k},Z',\alpha'}
\end{equation}
which is the form used in the main text. We insert \equref{FourierTransformOperators} also into the interlayer tunneling $\mathcal{H}_{\text{inter}}$,
\begin{align}
    \mathcal{H}_{\text{inter}} &= \frac{1}{\sqrt{N_1 N_2}}\sum_{\vec{R}_\ell \in \text{BL}_\ell} \sum_{\alpha_\ell=1}^{N_b^\ell} a^\dagger_{1,\vec{R}_1,Z=0,\alpha_1} t_{\alpha_1,\alpha_2}(\mathcal{R}_{\theta_1}\vec{R}_1 - \mathcal{R}_{\theta_2}\vec{R}_2)  a^\pdagger_{2,\vec{R}_2,Z=d_2,\alpha_2} + \text{H.c.}, \\
    &= \frac{1}{N_1 N_2} \sum_{\vec{k}_\ell \in \text{BZ}_{\ell,\theta_\ell}} \sum_{\vec{R}_\ell \in \text{BL}_{\ell,\theta_\ell}} \sum_{\alpha_\ell=1}^{N_b^\ell} e^{i(\vec{k}_2\vec{R}_2 - \vec{k}_1\vec{R}_1)} \bar{c}^\dagger_{1,\mathcal{R}_{-\theta_1}\vec{k}_1,Z=0,\alpha_1} t_{\alpha_1,\alpha_2}(\vec{R}_1 - \vec{R}_2)  \bar{c}^\pdagger_{2,\mathcal{R}_{-\theta_2}\vec{k}_2,Z=d_2,\alpha_2} + \text{H.c.}.
\end{align}
We observe, again, that the tunneling term is most naturally represented in the lab-frame operators (\ref{SecondTransformationLabFrame}) and introduce the (now matrix-valued) Fourier decomposition $t_{\alpha_1,\alpha_2}(\vec{R}) =  \sum_{\vec{q}} (T_{\vec{q}})_{\alpha_1,\alpha_2} e^{i\vec{q}\vec{R}}$. Using \equref{RLIdentity}, we arrive at the final form of the tunneling term
\begin{subequations}\begin{align}
    \mathcal{H}_{\text{inter}} &= \sum_{\vec{k}_{\ell}\in\text{BZ}_{\ell,\theta_\ell} } \sum_{\vec{q}} \sum_{\alpha_\ell=1}^{N_b^\ell} \delta_{\vec{k}_1-\vec{q} \in \text{RL}_{1,\theta_1}} \delta_{\vec{k}_2-\vec{q} \in \text{RL}_{2,\theta_2}} c^\dagger_{1,\vec{k}_1,Z=0,\alpha_1} (T_{\vec{q}})_{\alpha_1,\alpha_2} c^\pdagger_{2,\vec{k}_2,Z=d_2,\alpha_2}  + \text{H.c.} \\
    &= \sum_{\vec{k}_{\ell}\in\text{BZ}_{\ell,\theta_\ell} } \sum_{\vec{G}_\ell \in \text{RL}_{\ell,\theta_\ell}} \sum_{\alpha_\ell=1}^{N_b^\ell} \delta_{\vec{k}_1 +\vec{G}_1,\vec{k}_2 +\vec{G}_2} c^\dagger_{1,\vec{k}_1,Z=0,\alpha_1} (T_{\vec{k}_1 +\vec{G}_1})_{\alpha_1,\alpha_2} c^\pdagger_{2,\vec{k}_2,Z=d_2,\alpha_2}  + \text{H.c.}, 
\end{align}
\label{FormOfHinter}\end{subequations}
which is the analogue of \equref{FormOfH2P} and the form of the momentum-space interface Hamiltonian used in the main text. 

\subsection{Projection to surface states of topological insulators}\label{sec:app_projection_TI_surface}
To formally connect this Hamiltonian to the discussion of the topological insulator surface states, let us denote the eigenstates of $h_{\alpha,\alpha'}^\ell(\mathcal{R}_{-\theta_\ell}\vec{k},Z,Z')$ by $\phi_{\vec{k},n}$, i.e.,
\begin{equation}
    \sum_{\alpha',Z' \in \mathcal{Z}_\ell }h_{\alpha,\alpha'}^\ell(\mathcal{R}_{-\theta_\ell}\vec{k},Z,Z') (\phi_{\vec{k},n})_{Z',\alpha'} = \epsilon_{\vec{k},n} (\phi_{\vec{k},n})_{Z,\alpha}. \label{EigenvaluesOfBlochHam}
\end{equation}
If there is a finite bulk gap but in-gap surface states, associated with indices $n \in \mathcal{N}_\ell$ and around momenta $\vec{k} \in P_{\ell}$ in \equref{EigenvaluesOfBlochHam}, we can project at low energies to these modes. Instead of using $\phi_{\vec{k},n}$ directly, we perform this projection in a more general basis spanned by the unitarily related states
\begin{equation}
    \widetilde{\phi}_{\vec{k},\sigma} = \sum_{n\in \mathcal{N}_\ell} U_{\sigma,n}(\vec{k}) \phi_{\vec{k},n}, \qquad \sigma = 1,2 \dots, |\mathcal{N}_\ell|, \quad \vec{k}\in P_{\ell},
\end{equation}
where $U(\vec{k})$ is a unitary $|\mathcal{N}_\ell| \times |\mathcal{N}_\ell|$ matrix. We will only assume that $U(\vec{k})$ is chosen such that $\widetilde{\phi}_{\vec{k},\sigma}$ are smooth in $P_{\ell}$. For instance, for the single Dirac cone surface states analyzed in the main text, we have $|\mathcal{N}_\ell|=2$ and $\{\widetilde{\phi}_{\vec{k},\sigma}|\sigma=\uparrow, \downarrow\}$ are the wave functions transforming the same way as the spin quantum number (relating the ``pseudospin'' quantum number to the microscopic degrees of freedom). 

To project the Hamiltonian, we substitute
\begin{align}
    c_{\ell,\vec{k},Z,\alpha} \, \longrightarrow \,  \begin{cases} \sum_{\sigma=1}^{|\mathcal{N}_\ell|} \left(\widetilde{\phi}_{\vec{k},\sigma}\right)_{Z,\alpha} d_{\ell,\vec{k},\sigma} , \quad &\vec{k}\in P_\ell, \\
    0, \qquad &\vec{k}\notin P_\ell, \end{cases}\label{SurfaceStateProjection}
\end{align}
for the $\ell$ (both or only one of them) with in-gap surface states. In \equref{SurfaceStateProjection}, $d_{\ell,\vec{k},\sigma}$ are the electronic operators associated with these surface modes. For instance, if both materials are topological insulators, the projection, $\mathcal{H}_{\text{bulk}} + \mathcal{H}_{\text{inter}} \rightarrow \mathcal{H}^{\text{SS}}_{\text{bulk}} + \mathcal{H}^{\text{SS}}_{\text{inter}}$, leads to an effectively 2D theory with
\begin{equation}
    \mathcal{H}^{\text{SP}}_{\text{bulk}} = \sum_{\ell =1 ,2} \sum_{\vec{k}\in P_{\ell}}  \sum_{\sigma,\sigma' = 1}^{|\mathcal{N}_\ell|} d^\dagger_{\ell,\vec{k},\sigma} h_{\sigma,\sigma'}^{\text{SS},\ell}(\vec{k}) d^\pdagger_{\ell,\vec{k},\sigma'},
\end{equation}
where we introduced $h_{\sigma,\sigma'}^{\text{SS},\ell}(\vec{k}) := \sum_{Z,Z' \in \mathcal{Z}_\ell} \sum_{\alpha,\alpha'} \left(\widetilde{\phi}^*_{\vec{k},\sigma}\right)_{Z,\alpha} h_{\alpha,\alpha'}^\ell(\vec{k},Z,Z') \left(\widetilde{\phi}_{\vec{k},\sigma'}\right)_{Z',\alpha'}$. Furthermore, the interlayer coupling in \equref{FormOfHinter} becomes
\begin{equation}
    \mathcal{H}^{\text{SS}}_{\text{inter}} = \sum_{\vec{k}_{\ell}\in P_{\ell} } \sum_{\vec{q}} \sum_{\sigma_\ell = 1}^{|\mathcal{N}_\ell|}  \delta_{\vec{k}_1-\vec{q} \in \text{RL}_{1,\theta_1}} \delta_{\vec{k}_2-\vec{q} \in \text{RL}_{2,\theta_2}} d^\dagger_{1,\vec{k}_1,\sigma_1} (T^{\text{SS}}_{\vec{q}})_{\sigma_1,\sigma_2} d^\pdagger_{2,\vec{k}_2,\sigma_2}  + \text{H.c.}.
\end{equation}
Similarly, we here defined the projected tunneling Hamiltonian
\begin{equation}
    (T^{\text{SS}}_{\vec{q}})_{\sigma_1,\sigma_2}:= \sum_{\alpha_1,\alpha_2} \left(\widetilde{\phi}^*_{\vec{k}_1,\sigma_1}\right)_{Z=0,\alpha_1}(T_{\vec{q}})_{\alpha_1,\alpha_2} \left(\widetilde{\phi}_{\vec{k}_2,\sigma_2}\right)_{Z=d_2,\alpha_2}.
\end{equation}

\section{Additional subleading crystalline reference points for triangular and square lattices}\label{ap:more_shells}
To find the crystalline reference points and their momentum $Q$, we consider  a finite set of points from the square and triangular reciprocal lattices. For any such finite set, one will miss continuum reference points since the required reciprocal lattice vectors are not included. However, for the purpose of determining the leading (and next-to-leading etc.) continuum reference points, it is sufficient to consider a finite set only, as we explain next.

From the definition of $Q$,
\begin{equation}
    Q(\vec{G}_\ell,j_\ell) := \text{max}_{\ell=1,2} |\vec{P}_{j_\ell,\ell} + \vec{G}_\ell|,
\end{equation}
we can see that $Q$ increases for large values of $\vec{G}_{1,2}$. As such, once we have identified the leading continuum reference point using a sufficiently large number of shells, further increasing the number of shells cannot yield an additional continuum reference point with smaller momentum.

\begin{figure}[H]
    \centering
    \includegraphics[scale=0.9]{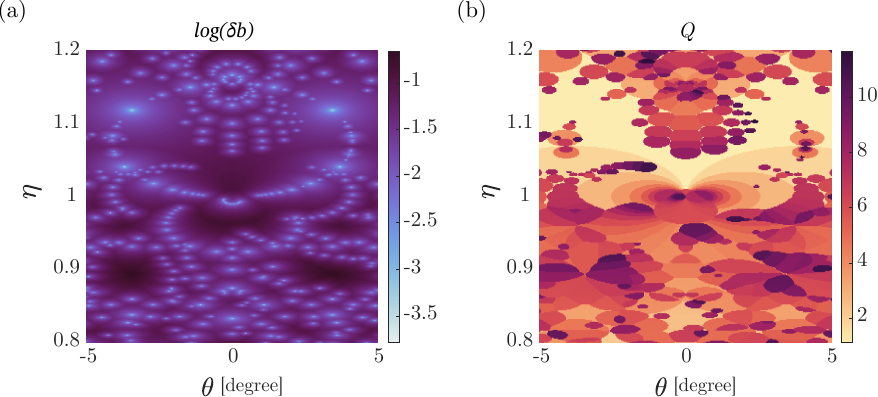}
    \caption{Scan for crystalline reference points for $(X_1,\Gamma)$. Here we considered 8 shells of lattice points. (a) shows the minimal $\delta b$ found for each combination of lattice distortions. (b) shows the respective $Q$.}
    \label{fig:more_shells}
\end{figure}

\begin{figure}[H]
    \centering
    \includegraphics[scale=0.9]{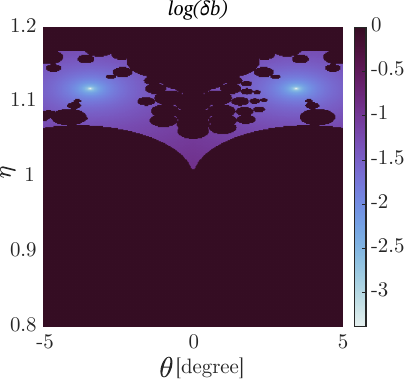}
    \caption{Minimal $\delta b$ for the $(X_1,\Gamma)$ case masked by the regions where $Q$ is also minimal.}
    \label{fig:masked_db}
\end{figure}

To exemplify this reasoning, we show the reference points for the $(X_1,\Gamma)$ case when we consider 8 shells of lattice points in \figref{fig:more_shells}. We can see that the number of reference points grows immensely in comparison to what we observe in \figref{fig:q_scan}(a). In order to distinguish leading and subleading order points, we create a mask for $\delta b$ with the lowest $Q$ value. In \figref{fig:masked_db}, we show the masked plot. Hence we only see the regions of $\delta b$ corresponding to the lowest $Q$. It is no surprise that we find the exact same leading order reference points from the main text.

\section{Example away from a crystalline reference point}\label{AwayFromAContinuumReferencePoint}

\begin{figure}[H]
    \centering
    \includegraphics{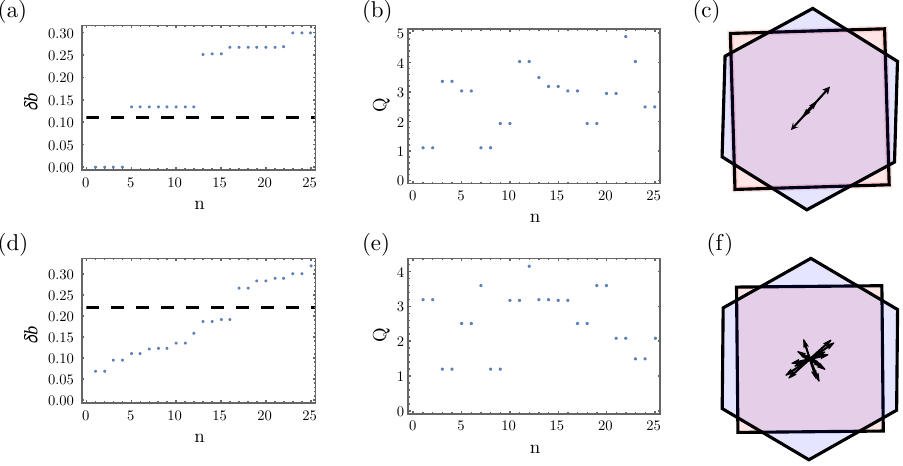}
    \caption{The first and second row of plots use a crystalline reference point and an arbitrary point respectively. In the first column (a) and (d), we show $|\delta \vec{b}|$ for a given $\vec{G}_{1,2}$ in ascending order. (b) and (e) show the respective $Q$ values. (c) and (f) exemplify $\delta \vec{b}$ found for small deviations in each case.}
    \label{fig:away_from_ref}
\end{figure}

In this appendix, we exemplify the qualitative difference between a crystalline reference point and an arbitrary point. One final time, we use the square/triangular lattice interface with the $(X_1,\Gamma)$ case as an example. For each combinations of $\vec{G}_{1,2}$, we calculate $\delta b$ and sort them in ascending order (label $n$). This is shown in \figref{fig:away_from_ref}(a) and (d). We also plot the respective $Q$ in (b) and (e), using the same label $n$.

In the first row, i.e., \figref{fig:away_from_ref}(a), (b) and (c), we consider a crystalline reference point. We can observe that $\delta b$ goes to 0 and has a rather small $Q$. Furthermore we note a clear hierarchy in $\delta b$, with a gap between smallest and subleading values. In (c), we plot $\delta \vec{b}$ arising from a small deviation and considering the vectors under the dashed line in (a). We can see 4 collinear and commensurate (the small vectors are 3 time smaller than the bigger ones) vectors. Hence, close to a reference point, there is a clear cutoff for which points to consider and which lead to a commensurate modulation.

On the other hand, in the second row of plots, i.e., \figref{fig:away_from_ref} (d), (e) and (f), we do not find $\delta b \rightarrow 0$ for any combination of $\vec{G}_{1,2}$. Neither do we find a clear hierarchy for $\delta b$. Moreover, the lowest $\delta b$ have high $Q$; the large but finite value of $Q$ for $n=1$ is related to the number of shells we consider here. We can see that, under these conditions, we end up with a non-commensurate and non-collinear set of $\delta b$, i.e., there is no well-defined crystalline moir\'e modulation.

\section{First order expansion for $\delta \vec{b}$}

Here we show the explicit form of $\delta \vec{b}$, considering the first-order expansion in angle and lattice mismatch around the leading order crystalline reference point $(\theta^*,\eta^*)$
\begin{equation}
\begin{split}
    &\delta \vec{b} (\theta,\eta) = \eta \,\,\mathcal{R}_{-\theta / 2} (\vec{P}_{j_2,2} + \vec{G}_2) - \mathcal{R}_{\theta/2}(\vec{P}_{j_1,1}+\vec{G}_1)\\
     &\delta \vec{b} (\theta,\eta) \sim \delta \vec{b} (\theta^*,\eta^*) + + (\eta - \eta^*) \,\,\frac{\partial\, \delta \vec{b}}{\partial \eta} \biggr\rvert_{(\theta^*,\eta^*)}+ (\theta - \theta^*) \,\,\frac{\partial\, \delta \vec{b}}{\partial \theta} \biggr\rvert_{(\theta^*,\eta^*)}.
    \end{split}
\end{equation}
For small deviations, we define $(\eta - \eta^*) \equiv \epsilon_\eta$ and $(\theta - \theta^*) \equiv \epsilon_\theta$. We summarize the results in Table \ref{tab:delta_b}. We note that in the cases where $C_{2z}$ is present, $-\delta \vec{b}$ is also valid. In the cases in which two valleys are related through $C_{2z}$, the $\delta \vec{b}$ at a given valley is also related to the other one through this symmetry.

\begin{table*}
\begin{center}
\caption{First order expansion of $\delta \vec{b}$ for small deviations $\epsilon_\theta$ and $\epsilon_\eta$ from the leading order crystalline reference point $(\theta^*,\eta^*)$.}
\label{tab:delta_b}
\begin{ruledtabular}
 \begin{tabular}{ccc} 
\multicolumn{2}{c}{high-symmetry point} & $\delta\vec{b}$ 
\\ \cline{1-2}  
Square & Triangular & 
\\ \hline
$\Gamma$ & $\Gamma$ &  $\vec{0}$, $(\epsilon_\eta,-\epsilon_\theta)$  \\
$\Gamma$ & $M$ & $(\epsilon_\theta,\frac{\sqrt{3}}{2}\epsilon_{\eta})$  \\
$\Gamma$ & $K$  & $(\epsilon_\theta,\frac{2}{\sqrt{3}}\epsilon_{\eta})$   \\ \hline
$M$ & $\Gamma$  & $(1.94 \epsilon_\eta - 1.44 \epsilon_\theta, -1.8\epsilon_\eta - 1.55 \epsilon_\theta )$ \\
$M$ & $M_1$ & $ ( \frac{\epsilon_\eta}{2} + \frac{3}{2} \epsilon_\theta, -\frac{\epsilon_\theta}{2} + \frac{3}{2} \epsilon_\eta ) $   \\
 & $M_{3(2)}$  & $( (-)1.73 \epsilon_\eta - 0.44 \epsilon_\theta, -0.5 \epsilon_\eta -(+) 1.52 \epsilon_\theta )$  \\
 & $M_{3(2)}$  & $( 1.25 \epsilon_\eta +(-)0.51 \epsilon_\theta, +(-)0.43 \epsilon_\eta -1.5 \epsilon_\theta )$ \\
$M$ & $K$ & $(0.49 \epsilon_\eta - 1.5 \epsilon_\theta, -1.45\epsilon_\eta - 0.51 \epsilon_\theta )$  \\ \hline
$X$ & $\Gamma$ & $(-0.47 \epsilon_\eta -0.98 \epsilon_\theta , -0.88 \epsilon_\theta + 0.53 \epsilon_\eta  )$  \\
$X$ & $M$ & $(2.5 \epsilon_\eta,-2.5\epsilon_\theta)$ \\
$X$ & $K$ & $(-0.5 \epsilon_\theta, -\frac{\sqrt{3}}{3} \epsilon_{\eta})$  \\
\end{tabular}
\end{ruledtabular}
\end{center}
\end{table*}

\section{Additional band structure plots}\label{sec:app_additioan_band_struc}
In this section, we calculate the band structure resulting from a TI ($\ell=1$) and a 3D NI ($\ell=2$) with surface Dirac cones and electron pockets around the $\Gamma$ points. The Hamiltonian reads as a combination of \equref{eq:hamiltonian_nini} and \equref{eq:hamiltonian_titi}. At the interface the $2\times 2$ pseudospin Hamiltonians are given by $h_{\vect{k}}^{\ell=1}=v_{\ell=1} \left[ \sigma_y k_x - \sigma_x k_y \right] + E_{\ell=1} \sigma_0$ and  $h_{\vect{k}}^{\ell=2}=\left(\vec{k}^2/2m^*_{\ell=2} - \mu_{\ell=2}\right)\sigma_0$. Away from the interface, we allow for vertical hopping along the positive $z$ direction to account for the 3D bulk structure of the NI. The full Hamiltonian is therefore given by
\begin{align}
    \mathcal{H}_{\text{tTI-NI}} =& \sum_{|\vec{k}| < \Lambda}\sum_{Z=0}^{N_Z-1} c^\dagger_{\ell,\vec{k},Z} \begin{pmatrix}h^{\ell = 1}_{R_{-\theta_1}\vec{k}}\delta_{Z,0} & T_{0}\delta_{Z,0}  \\ T^\dagger_{0}\delta_{Z,0}  & h^{\ell = 2}_{R_{-\theta_2}\vec{k}} \end{pmatrix}_{\ell,\ell'} \hspace{-0.5em} c^\pdagger_{\ell',\vec{k},Z}\\ \nonumber
    &+ \left[ \sum_{|\vec{k}| < \Lambda}\sum_{\pm} c^\dagger_{1,\vec{k},Z=0} T_\pm c^\pdagger_{2,\vec{k} \pm \delta\vec{b},Z=0} + \text{H.c.} \right]
    +t_z\left[ \sum_{|\vec{k}| < \Lambda}\sum_{Z=1}^{N_Z-1} c^\dagger_{\ell=2,\vec{k},Z+1} c^\pdagger_{\ell=2,\vec{k},Z} + \text{H.c.} \right].
\end{align}
The interface interaction terms $T_0$ and $T_\pm$ are defined in the same way as in \secref{sec:Expl_calc_TiTi}.

\begin{figure}
    \centering
    \includegraphics[width=0.5\linewidth]{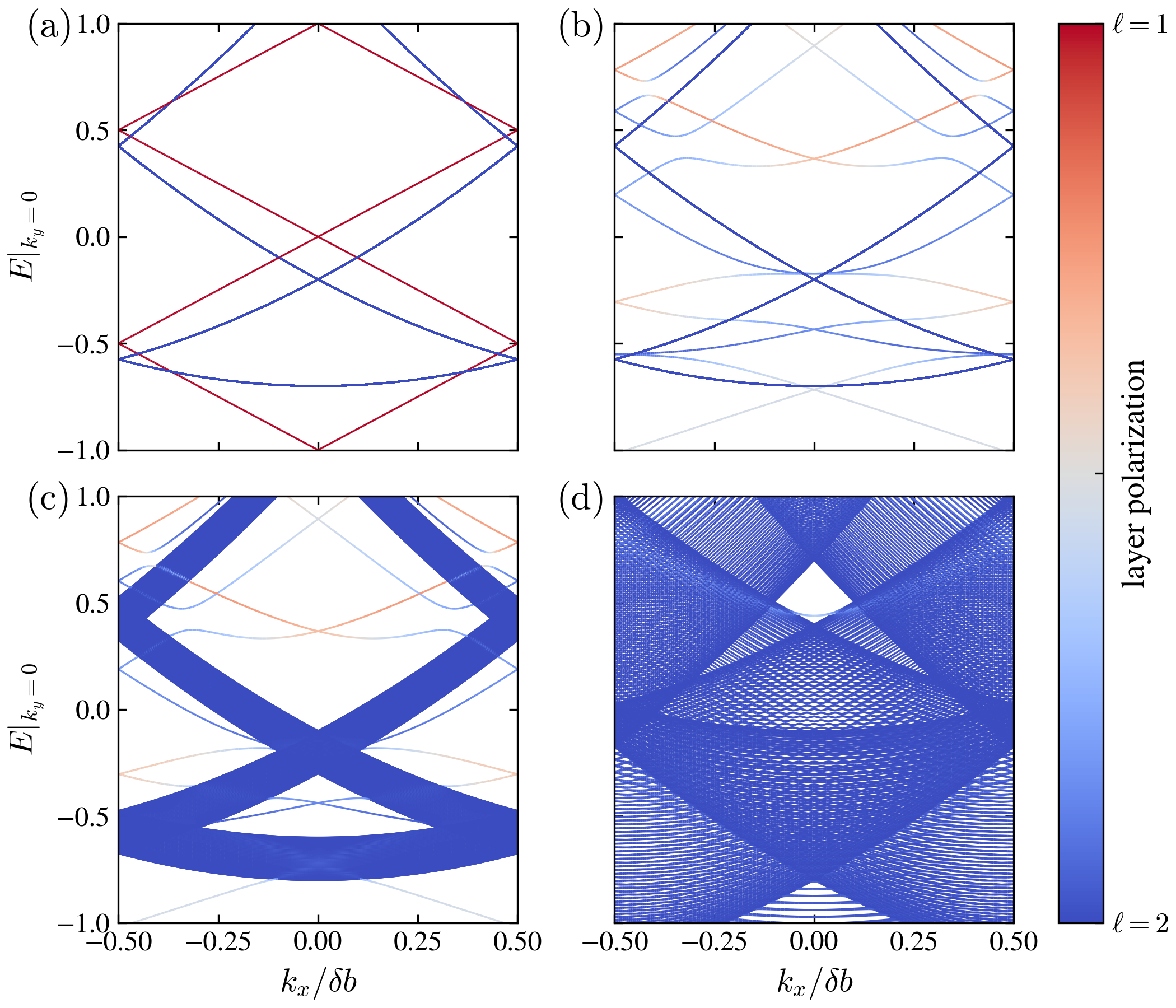}
    \caption{Band structure of 3D NI stacked on top of a 2D TI, with electron pockets at the $\Gamma$ point in both materials. Panel (a) shows bare band structures without any interface couplings. The effect of non-vanishing interface interaction terms $T_0$ and $T_\pm$ is shown in panel (b). In (c) and (d) we plot the same band structures with an additional a weak (c) and strong (d) vertical hopping term in the 3D NI. The color of the bands indicates the layer polarization of the respective wave functions.}
    \label{fig:TI-NI}
\end{figure}

Figure~\ref{fig:TI-NI}(a) shows the respective band structure without any interface couplings, \ie $T_0=T_\pm=0$ and $t_z=0$. Similar to the main text, we measure all energies in terms of the natural moir\'e energy scale $v_1\db$. The color indicates the layer polarization. In this edge case, we clearly see the folded Dirac cones (square bands) in the $\ell=1(2)$ layers. Figure~\ref{fig:TI-NI}(b) shows the system with non-zero interface interactions $T_0=0.3\sigma_0$, $T_\pm=0.3\sigma_0$ and no vertical hopping \ie $t_z=0$. We note that there are some avoided crossings, however as there is no clear pattern in which the respective bands intersect this does not to lead to a specific structure. The interface bands are show no clear layer polarization anymore. The $\ell=2$ polarized square bands originate from the degenerate band structure of the $N_Z-1$ non-interface layers of the 3D material. In \figref{fig:TI-NI}(c) the hopping $t_z=0.05$ is tuned up to a small value which corresponds to the limit of a layered material. Similar to \figref{fig:NiNi}(a) this splits up degenerate bulk bands. In between those bands there are still some interface-localized bands. Those are entirely swallowed by the bulk bands when we set $t_z=0.3$ in \figref{fig:TI-NI}(d). In this region the bulk dominates the whole band structure.
\end{appendix}

\end{document}